\documentclass[amsmath,trackchanges,twocolumn]{aastex702}

\begin{document}

\title{Weak Radio and Bright Early Optical Emission in EP260321a: A Comparative Study with XRO~080109 Reveals the Role of Circumstellar Material Mass}

\author[0000-0002-0170-0741]{Cui-Yuan Dai}
\affiliation{School of Astronomy and Space Science, Nanjing University, Nanjing 210023, China; xywang@nju.edu.cn}
\affiliation{Key Laboratory of Modern Astronomy and Astrophysics (Nanjing University), Ministry of Education, Nanjing
210023, China}
\email[]{cydai@smail.nju.edu.cn}

\author[0000-0002-5881-335X]{Xiang-Yu Wang}
\affiliation{School of Astronomy and Space Science, Nanjing University, Nanjing 210023, China; xywang@nju.edu.cn}
\affiliation{Key Laboratory of Modern Astronomy and Astrophysics (Nanjing University), Ministry of Education, Nanjing
210023, China}
\email[]{xywang@nju.edu.cn}  

\author[0000-0001-5751-633X]{Jian-He Zheng}
\affiliation{School of Astronomy and Space Science, Nanjing University, Nanjing 210023, China; xywang@nju.edu.cn}
\affiliation{Key Laboratory of Modern Astronomy and Astrophysics (Nanjing University), Ministry of Education, Nanjing
210023, China}
\email[]{mg21260020@smail.nju.edu.cn}

\begin{abstract}
Shock breakout (SBO)  occurs when
a radiation-mediated shock wave,  formed during a core-collapse  supernova (SN) explosion, breaks out from the stellar surface of the progenitor or from the circumstellar material (CSM), if the CSM is optically thick. 
The X-ray outburst XRO~080109, associated with a Type Ib/c SN~2008D, has been suggested to be an SBO event from a regular supernova. The recent detection of EP260321a, associated with a broad-lined Type Ic SN~2026gzf, by the {\em Einstein Probe} adds a new SBO event. However, the two SBO events differ markedly in their spectra, energetics, late-time radio emission and early optical emission within the first day. 
We show that the non-detections in radio follow-up observations of EP260321a imply a shock velocity $\lesssim0.1c$, whereas the radio data of XRO~080109 suggest a shock velocity of $\simeq0.2c$. These velocities are consistent with those inferred from the properties of the X-ray outbursts and  explain the difference in their X-ray  spectra.  
The X-ray energy and duration of EP260321a indicate that the shock breaks out from the  CSM  at a radius of $\sim2\times10^{13}\ {\rm cm}$. Its bright optical emission at $t\lesssim1\ {\rm d}$ can be explained by cooling emission from a dense, extended CSM shell heated by the radiation-mediated shock, requiring a total CSM mass of $\sim0.06M_\odot$, substantially larger than that inferred for XRO~080109. Finally, the lower shock velocity and larger CSM mass can be unified in a picture in which a more massive CSM decelerates the radiation-mediated shock more significantly before it breaks out. 
 
\end{abstract}

\section{Introduction}
The earliest electromagnetic signal from a core-collapse supernova (SN) is the shock breakout (SBO) \citep{Colgate1974,Falk1978,Klein1978,Ensman1992,Matzner1999,Katz2010,Piro2010,Nakar2010}.
During core collapse, a strong radiation-mediated shock propagates through the exploding star.
SBO occurs when the shock reaches the stellar edge and the optical depth ahead of it drops below $\tau\simeq c/v_{\rm sh}$, where $v_{\rm sh}$ is the shock velocity, producing a bright X-ray/UV flash on timescales of seconds to a fraction of an hour.
If sufficient circumstellar material (CSM) has been ejected before explosion, either through a steady wind or an episodic outer-envelope ejection, the breakout can instead occur at larger radii within the CSM, provided that its optical depth exceeds $c/v_{\rm sh}$.

The low-luminosity gamma-ray burst (GRB) 060218, associated with a broad-lined Type Ic supernova SN~2006aj, was the first event proposed as shock breakout from a Wolf--Rayet (WR) star embedded in a dense CSM \citep{Campana2006,Waxman2007,Wang07}. 
Its large radiated energy, $10^{49}$--$10^{50}\ {\rm erg}$, requires a mildly relativistic shock, $v_{\rm sh}\sim0.8c$, which is difficult to generate in a regular SN explosion but suggests that an additional central engine should play a role. The X-ray transient XRO~080109, associated with a Type Ib/c SN~2008D, was discovered serendipitously during a scheduled \textit{Swift} observation of the nearby galaxy NGC~2770 \citep{Soderberg2008}.
It has an  energy $E_X=2\times10^{46}\ {\rm erg}$ in  0.3--10 keV and a smooth light curve with a rise time of $80\ {\rm s}$ and a duration of about 400 s, consistent with shock breakout from a regular SN \citep{Soderberg2008,Modjaz2009}.
Several studies suggest that it originates from an exploding WR star  embedded in an optically-thick wind, producing a shock breakout with a velocity of $v_{\rm sh}\simeq0.2$--$0.3c$ \citep[e.g.,][]{Katz2010,Balberg2011,Svirski2014}.
Its 0.3--10 keV spectrum is non-thermal, with a power-law photon index $\Gamma_X=2.3\pm0.3$ \citep{Soderberg2008, Mazzali2008Sci...321.1185M}.
This non-thermal spectrum is expected for shocks with $v_{\rm sh}\gtrsim0.1c$, for which the shock transition is far from thermal equilibrium \citep{Katz2010,Nakar2010}.

{
The \textit{Einstein Probe} (EP) has recently discovered several extragalactic fast X-ray transients (FXTs) associated with SNe Ic-BL, revealing a diversity of high-energy emission from stripped-envelope stellar explosions. In the 0.5--4 keV band, EP240414a/SN~2024gsa and EP250108a/SN~2025kg radiated isotropic-equivalent X-ray energies of $E_{\rm X,iso}\simeq5\times10^{49}\ {\rm erg}$ and $\simeq(0.5\mbox{--}3)\times10^{49}\ {\rm erg}$, respectively \citep{Sun2025,Li2025}, more than two orders of magnitude larger than that of XRO~080109. EP240414a has been suggested to be produced by a weak relativistic jet, while jet--cocoon and off-axis jet scenarios have also been proposed \citep{Sun2025,vanDalen2025,Zheng2025ApJ...985...21Z,Bright2025}. For EP250108a, the proposed interpretations include a weak or trapped relativistic jet, jet--cocoon emission, and relativistic shock breakout associated with an engine-driven outflow \citep{Li2025,EylesFerris2025,Rastinejad2025,Srinivasaragavan2025}. EP250827b/SN~2025wkm is less energetic, with $E_{\rm X,iso}\sim10^{48}\ {\rm erg}$. Its X-ray emission and the unusually luminous first optical peak have been jointly explained by a central-engine-driven outflow, in which magnetically driven winds from a newborn magnetar and accretion disk break out from an extended CSM \citep{Srinivasaragavan2026}. This favors an engine-driven breakout rather than an SBO powered by a regular SN shock. Together,  these FXTs associated with  Ic-BL SNe are more naturally linked to relativistic or engine-driven outflows than to SBOs powered by regular SN shocks.}

More recently, EP260321a, the closest extragalactic FXT discovered by EP to date at $z=0.0343$, was found to show properties consistent with an SBO  from a regular SN \citep{Yuan2026}. It was followed by a broad-lined Type Ic supernova (SN Ic-BL), pointing to a Wolf--Rayet progenitor \citep{Yuan2026,Chen2026,Martin-Carrillo2026,OConnor2026,Rastinejad2026}.
The X-ray emission is well described by a blackbody with $kT=0.12\ {\rm keV}$.
The light curve is smooth, with a rise time of $690\pm60\ {\rm s}$ and a  duration of $T_{90}\sim 1750\,{\rm s}$.
The outburst reached a peak luminosity of $L_{{\rm p}, X}=(1.0\pm0.3)\times10^{45}\ {\rm erg\ s^{-1}}$ and radiated a total energy of $E_X=(8.7\pm0.6)\times10^{47}\ {\rm erg}$.
Together, the thermal spectrum, smooth temporal evolution, and extremely low luminosity support an SBO origin for EP260321a.
The long duration and large energy  favor  shock breakout from a CSM shell rather than from the progenitor surface \citep{Yuan2026}.

Although both XRO~080109 and EP260321a have been interpreted as shock breakout from regular SN explosions, their observed properties differ markedly.
First, the radio follow-up nondetections of EP260321a imply a radio luminosity at least an order of magnitude lower than that of XRO~080109, despite the fact that the SBO energy of EP260321a is larger by a factor of $\sim40$ \citep{Yuan2026,Rastinejad2026}.
Second, EP260321a shows   bright optical emission at $0.1$--$1$ day after the X-ray outburst \citep{Yuan2026,Chen2026,Martin-Carrillo2026}.
Its luminosity at $\sim 0.1 $ day is significantly higher than that of XRO~080109 \citep{Soderberg2008}.
Here we investigate the physical origin of the differences between these two SN shock-breakout events.

The paper is organized as follows. Sect.~\ref{sec:shock_velocity} investigates the constraints on the shock velocity of the two SBO events using the X-ray emission and radio data. Sect.~\ref{sec:csm_cooling} studies the cooling emission from the shocked CSM. Sect.~\ref{sec:unified_interpretation} presents a unified interpretation of the two events, and Sect.~\ref{sec:summary} summarizes our conclusions.

\section{Observational Inference of the Shock Velocity}
\label{sec:shock_velocity}

We constrain the shock velocity of XRO~080109 and EP260321a using two independent diagnostics.
The X-ray  emission constrains the velocity during the breakout through its energy, duration, and spectrum, whereas the late-time radio emission traces the fastest post-breakout ejecta.
As the ejecta interact with the wind at large radii, the shock transitions from a radiation-mediated shock to a collisionless shock, which  accelerates relativistic electrons and produces synchrotron radio emission. The intensity of the radio emission is sensitive to the initial velocity of the fastest ejecta, so the shock breakout velocity can be restricted with the follow-up radio emission.

\subsection{Constraints from the X-ray emission}
\label{subsec:xray_velocity}

The X-ray breakout emission provides a direct estimate of the shock velocity.
For both XRO~080109 and EP260321a, the observed X-ray energy and duration suggest shock breakout in optically thick CSM outside the stellar surface.
Radiation begins to escape when the shock reaches the breakout radius $R_{\rm bo}$, where the diffusion time becomes comparable to the expansion time, or equivalently when the optical depth ahead of the shock satisfies
$\tau_{\rm bo}\simeq c/v_{\rm bo}$.
Here, $v_{\rm bo}$ is the shock velocity at the breakout time.
For a stellar wind with density profile
$\rho=A r^{-2}$, where $A$ is the wind-density normalization
($A=5\times10^{11}A_*\,{\rm g\,cm^{-1}}$),
and outer radius $R_{\rm w}$, the optical depth of the material ahead of the shock at radius $r$ is
$\tau_{\rm w}=(r^{-1}-R_{\rm w}^{-1})\kappa A$,
where $\kappa$ is the opacity. The breakout condition can therefore be written as
$R_{\rm d}^{-1}=R_{\rm bo}^{-1}-R_{\rm w}^{-1}$, where
$R_{\rm d} \equiv \kappa A v_{\rm bo}/c$.
The characteristic radius $R_{\rm d}$ corresponds to the breakout radius for an infinitely extended wind \citep{Chevalier2011}.
Following \citet{Chevalier2011,Haynie2021}, we distinguish between the compact-wind regime ($R_{\rm w}<R_{\rm d}$) and the extended-wind regime ($R_{\rm w}>R_{\rm d}$).

The shock-breakout condition is
\begin{equation}
\label{Eq_tau_bo}
\tau_{\rm w,bo}
\simeq
\kappa\rho_0\Delta R_{\rm bo}
\simeq
\frac{c}{v_{\rm bo}}.
\end{equation}
{We define the SBO region as the radial interval from which photons can effectively escape on the dynamical timescale during the breakout,
$R_{\rm bo}\leq r\leq R_{\rm bo}+\Delta R_{\rm bo}$.
Its width is
$\Delta R_{\rm bo}\simeq v_{\rm bo}t_{\rm diff}$,
where $t_{\rm diff}$ is the photon diffusion time across this  region and $\rho_0$ is the characteristic upstream density in this region.}

\begin{figure*}[htbp]
\centering
\includegraphics[width=1.0\linewidth]{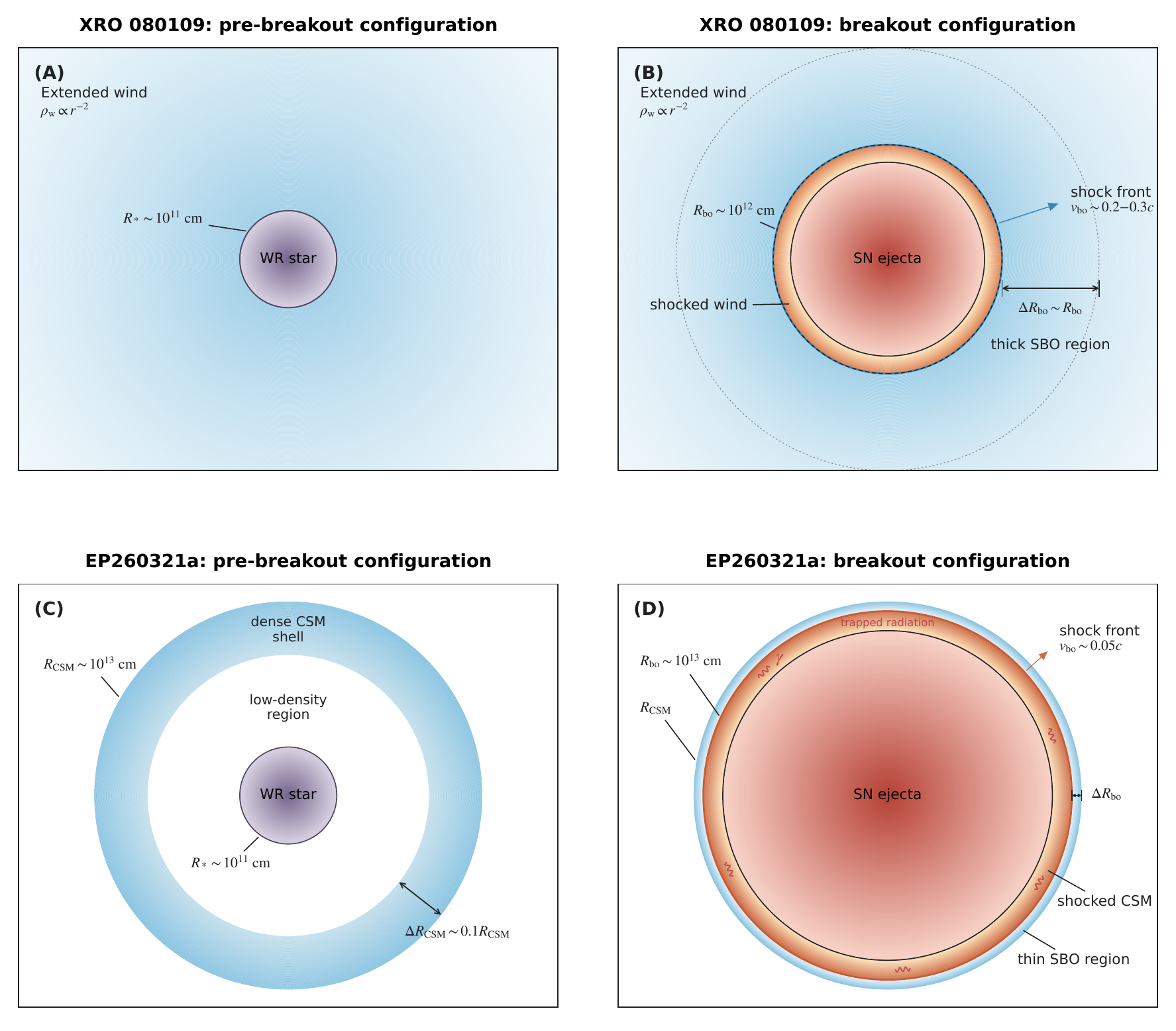}
\caption{\textbf{Pre-breakout and breakout configurations.}
Schematic comparison of the environments inferred for XRO~080109/SN~2008D and EP260321a/SN~2026gzf. In all panels, the blue region denotes the unshocked CSM, while the region between the SN ejecta and the forward shock denotes the shocked CSM. Panels (A) and (B) show the XRO~080109 case, in which the SN ejecta expand directly into an extended Wolf--Rayet wind and shock breakout occurs within the wind. Panel (B) depicts the onset of breakout, when the shock enters the SBO region. The breakout phase ends after the shock has traversed its radial extent, $\Delta R_{\rm bo}$. The breakout shock is relatively fast, $v_{\rm bo}\sim0.2$--$0.3c$, and the SBO region is geometrically thick, with $\Delta R_{\rm bo}\sim R_{\rm bo}$. Panels (C) and (D) show the EP260321a case, in which the SN ejecta first propagate through a low-density region before encountering a compact, dense CSM shell. Breakout occurs near the outer edge of the shell, at $R_{\rm bo}\simeq R_{\rm CSM}\sim10^{13}\ {\rm cm}$. The shell has a radial width of $\Delta R_{\rm CSM}\sim0.1R_{\rm CSM}$, as inferred from the optical cooling emission (Sect.~\ref{sec:csm_cooling}). Panel (D) likewise depicts the onset of breakout, when the shock enters the thin outer SBO region. The breakout velocity is much lower, $v_{\rm bo}\sim0.05c$. In this case, $\Delta R_{\rm bo}\ll\Delta R_{\rm CSM}$, so only a thin outer portion of the CSM shell contributes to the prompt breakout emission, whereas radiation deposited deeper in the shocked CSM remains trapped and emerges later as cooling emission.
}
\label{fig_shock_breakout_schematic}
\end{figure*}

Previous studies have shown that XRO~080109 is well described by an SBO in the extended-wind regime \citep{Balberg2011,Svirski2014}, for which
$R_{\rm bo}\simeq R_{\rm d}$. In this case, we have $\Delta R_{\rm bo}\simeq c/(v_{\rm bo}\kappa \rho_0 )\simeq R_d\simeq R_{\rm bo}$ (see Fig.~\ref{fig_shock_breakout_schematic} for a schematic plot).
Approximating the radiated breakout energy as the kinetic energy of the SBO region gives the following equation:
\begin{equation}
\label{Eq_E_bo}
E_{\rm bo}
\simeq
\frac{1}{2}M_{\rm bo}v_{\rm bo}^{2}
\simeq
2\pi R_{\rm bo}^{2}
\Delta R_{\rm bo}
\rho_0v_{\rm bo}^{2},
\end{equation}
where
$M_{\rm bo}\simeq4\pi R_{\rm bo}^{2}\Delta R_{\rm bo}\rho_0$.
Correction factors of order unity associated with the detailed shock structure  and breakout geometry are neglected.

In the extended-wind regime, the breakout radius is set by the shock-crossing time,
\begin{equation}
\label{Eq_R_bo}
R_{\rm bo}
\simeq
\beta_{\rm bo}ct_{\rm bo},
\end{equation}
{where $\beta_{\rm bo}\equiv v_{\rm bo}/c$, and $t_{\rm bo}$ denotes the breakout time, which is comparable to the observed  rise time of the pulse.} Combining Eqs.~\ref{Eq_tau_bo}--\ref{Eq_R_bo} gives
\begin{align}
\label{Eq_v_bo}
v_{\rm bo}
&=
\left(
\frac{\kappa E_{\rm bo}}
{2\pi ct_{\rm bo}^{2}}
\right)^{1/3}
\notag\\
&=
1.18\times10^{10}
\kappa_{0.2}^{1/3}
E_{\rm bo,46}^{1/3}
t_{\rm bo,80}^{-2/3}
\ {\rm cm\ s^{-1}},
\end{align}
or
\begin{equation}
\label{Eq_beta_bo}
\beta_{\rm bo}
=
0.39\,
\kappa_{0.2}^{1/3}
E_{\rm bo,46}^{1/3}
t_{\rm bo,80}^{-2/3}.
\end{equation}
Here,
$E_{\rm bo,46}\equiv E_{\rm bo}/10^{46}\,{\rm erg}$,
$t_{\rm bo,80}\equiv t_{\rm bo}/80\,{\rm s}$, and
$\kappa_{0.2}\equiv\kappa/(0.2\,{\rm cm^2\,g^{-1}})$. Thus,  the X-ray breakout of XRO~080109/SN~2008D requires a fast shock with
$\beta_{\rm bo}\sim0.4$.
The corresponding breakout radius is
\begin{equation}
\label{Rbo_wind}
R_{\rm bo}
\simeq
\beta_{\rm bo}ct_{\rm bo}
=
10^{12}\,
\beta_{\rm bo,0.4}
t_{\rm bo,80}
\ {\rm cm}.
\end{equation}

We next examine whether the extended-wind interpretation can also apply to EP260321a.
In the extended-wind regime, $R_{\rm bo}\simeq R_{\rm d}$, so the characteristic breakout duration is
\begin{equation}
t_{\rm bo}
\simeq
\frac{R_{\rm d}}{v_{\rm bo}}
=
\frac{\kappa A}{c}
\end{equation}
\citep{Ofek2010,Balberg2011,Chevalier2011,Haynie2021}.
For a radiation-dominated shock, the equilibrium blackbody temperature at breakout is set by the post-shock radiation pressure,
\begin{equation}
\label{Eq_int_energy}
\frac{aT_{\rm BB,bo}^{4}}{3}
\simeq
\frac{6}{7}\rho_{\rm 0}v_{\rm bo}^{2}.
\end{equation}
Combining Eq.~\ref{Eq_int_energy} with the breakout condition,
$\kappa \rho_{\rm 0} \Delta R_{\rm bo}=c/v_{\rm bo}$,
and $\Delta R_{\rm bo}\simeq v_{\rm bo}t_{\rm bo}$ gives (see also \cite{Chevalier2011})
\begin{equation}
kT_{\rm BB, bo}
\simeq
45\,{\rm eV}\,
\kappa_{0.2}^{-1/4}
t_{\rm bo,690}^{-1/4}.
\end{equation} This temperature is approximately a factor of $3$ below the observed value,
$kT_{\rm obs}=124^{+7}_{-6}\,{\rm eV}$ \citep{Yuan2026}.
The extended-wind solution also requires
\begin{equation}
\beta_{\rm bo}
=
0.44\,
\kappa_{0.2}^{1/3}
E_{\rm bo,48}^{1/3}
t_{\rm bo,690}^{-2/3},
\end{equation}
where
$E_{\rm bo,48}\equiv E_{\rm bo}/10^{48}\,{\rm erg}$.
{Such a fast radiation-mediated shock is expected to be far from local thermal equilibrium, making it difficult to reproduce the observed  blackbody-like X-ray spectrum \citep{Katz2010,Nakar2010}. These arguments therefore disfavor an extended-wind breakout interpretation for EP260321a.}

We then consider an alternative scenario in which the shock breaks out near the outer edge of a compact wind, such that $R_{\rm bo}\simeq R_{\rm w}=R_{\rm CSM}$
(Fig.~\ref{fig_shock_breakout_schematic}). If the observed duration is set primarily by the light-crossing time $t_{\rm lc}$, the breakout radius is
\begin{equation}
R_{\rm bo}
\simeq
ct_{\rm lc}
=
2.1\times10^{13}
t_{\rm lc,690}
\ {\rm cm}.
\end{equation}
Combining the breakout condition,
$\kappa\rho_0\Delta R_{\rm bo}\simeq c/v_{\rm bo}$,
with Eq.~\ref{Eq_E_bo} gives
\begin{align}
\label{Eq_v_bo_260321}
v_{\rm bo}
&=
\frac{\kappa E_{\rm bo}}
{2f_{\rm sh} \pi cR_{\rm bo}^{2}}
\notag\\
&=
2.4\times10^{9} f_{\rm sh}  ^{-1}
\kappa_{0.2}
E_{\rm bo,48}
R_{\rm bo,2.1}^{-2}
\ {\rm cm\ s^{-1}},
\end{align}
or
\begin{equation}
\label{Eq_beta_bo_260321}
\beta_{\rm bo}
=
0.08\,
f_{\rm sh} ^{-1} \kappa_{0.2}
E_{\rm bo,48}
R_{\rm bo,2.1}^{-2},
\end{equation}
where $R_{\rm bo,2.1}
\equiv
{R_{\rm bo}}/
({2.1\times10^{13}\ {\rm cm}})$.
{Here we introduce a dimensionless factor $f_{\rm sh} \sim\mathcal{O}(1)$ to account for the uncertainty in the calculation of the breakout energy, which depends on the detailed shock structure and the breakout geometry \citep[e.g.,][]{Balberg2011,Matzner1999,Haynie2021}.} Taking a slightly larger value of $f_{\rm sh}$ or $R_{\rm bo}$ ($R_{\rm bo}=2.7\times10^{13}\ {\rm cm}$) would give $\beta_{\rm bo}\simeq0.05$, close to the thermal-equilibrium boundary discussed below. We therefore consider two cases in the following discussions: a thermal equilibrium case where the radiation spectrum is a blackbody, 
and a photon-starved case in which the radiation and gas are out of thermal equilibrium while the spectrum can still appear Wien-like and remains consistent with the observed blackbody-like X-ray spectrum.

\paragraph{\bf (1)  local thermal equilibrium case: $\beta_{\rm bo}\lesssim 0.05$.}

We first examine whether a shock with
$\beta_{\rm bo}\simeq0.05$ can produce the observed soft X-ray spectrum in approximate thermal equilibrium.
For a radiation-dominated shock, the pressure equilibrium condition gives (Eq.~\ref{Eq_int_energy})
\begin{align}
\rho_0
&=
\frac{7aT_{\rm BB, bo}^{4}}
{18v_{\rm bo}^{2}}
\notag\\
&=
4.9\times10^{-9}
\beta_{\rm bo,0.05}^{-2}
\left(
\frac{kT_{\rm BB, bo}}
{0.12\,{\rm keV}}
\right)^{4}
\ {\rm g\ cm^{-3}}.
\label{Eq_rho0_compact}
\end{align}
The optical-depth condition then determines the radial width of the SBO region:
\begin{align}
\Delta R_{\rm bo}
&=
\frac{c}
{\kappa\rho_0v_{\rm bo}}
\notag\\
&=
2.0\times10^{10}
\kappa_{0.2}^{-1}
\beta_{\rm bo,0.05}
\left(
\frac{kT_{\rm BB, bo}}
{0.12\,{\rm keV}}
\right)^{-4}
\ {\rm cm}.
\label{Eq_DeltaR_compact}
\end{align}
Relative to the adopted breakout radius,
\begin{equation}
\frac{\Delta R_{\rm bo}}
{R_{\rm bo}}
=
7.6\times10^{-4}
\kappa_{0.2}^{-1}
\beta_{\rm bo,0.05}
R_{\rm bo,2.66}^{-1}
\left(
\frac{kT_{\rm BB, bo}}
{0.12\,{\rm keV}}
\right)^{-4}.
\end{equation}
The SBO region is therefore geometrically thin. We note that this layer represents only the outermost portion of the CSM involved in the X-ray breakout, rather than its full radial extent. The deeper CSM structure is instead constrained by the subsequent cooling emission, which requires a detached shell with $\Delta R_{\rm CSM}\sim0.1R_{\rm bo}$ to reproduce the early optical emission of EP260321a (see Sect.~\ref{sec:csm_cooling}).

The corresponding photon diffusion time across the SBO region is
\begin{align}
t_{\rm diff}
&\simeq
\frac{\tau_{\rm bo}\Delta R_{\rm bo}}{c}
=
\frac{\Delta R_{\rm bo}}{v_{\rm bo}}
\notag\\
&=
13.5\,
\kappa_{0.2}^{-1}
\left(
\frac{kT_{\rm BB, bo}}
{0.12\,{\rm keV}}
\right)^{-4}
\ {\rm s}.
\label{Eq_tdiff_compact}
\end{align}
This diffusion time is much shorter than the observed rise time, 
$690\ {\rm s}$, consistent with our assumption that the rise time is determined by the light-crossing time.

To determine whether free--free emission can produce enough photons to establish thermal equilibrium, we use the thermal-coupling parameter defined by \citet{Nakar2010},
\begin{align}
\label{Eq_eta_th}
\eta_{\rm th}
&\equiv
\frac{n_{\rm BB}}
{\dot{n}_{\rm ph,ff}(T_{\rm BB})t_{\rm diff}} \notag \\
&\simeq
\frac{7\times10^{5}\ {\rm s}}
{t_{\rm diff}}
\left(
\frac{\rho}
{10^{-10}\ {\rm g\ cm^{-3}}}
\right)^{-2}
\left(
\frac{kT_{\rm BB}}
{100\ {\rm eV}}
\right)^{7/2} \notag \\
&\simeq
0.82\,
\kappa_{0.2}
\beta_{0.05}^{4}
\left(
\frac{kT_{\rm BB}}
{0.12\,{\rm keV}}
\right)^{-1/2},
\end{align}
where $n_{\rm BB}$ is the blackbody photon number density and
$\dot{n}_{\rm ph,ff}$ is the free--free photon production rate.
The final step is evaluated at the post-shock density,
$\rho_{\rm ps}\simeq\chi\rho_0$, with compression ratio $\chi\simeq7$.
The thermal equilibrium boundary, defined by
$\eta_{\rm th,bo}=1$, corresponds to
$\beta_{\rm bo}\simeq0.052$. Therefore, shocks with $\beta_{\rm bo}\lesssim 0.05$ lie below this boundary,
allowing the shocked material to reach thermal equilibrium.

\paragraph{\bf (2) Photon-starved solution: $\beta_{\rm bo}\gtrsim 0.05$.}

We next consider an alternative case with
$\beta_{\rm bo}=0.08$ as an example.
For this higher velocity, free--free emission cannot produce enough photons to maintain thermal equilibrium, so the radiation becomes photon-starved and the observed color temperature $T_{\rm obs}$ exceeds the equilibrium blackbody temperature $T_{\rm BB, bo}$.
Following \citet{Nakar2010},
\begin{equation}
T_{\rm obs}
=
T_{\rm BB, bo}
\frac{\eta_{\rm th,bo}^{2}}
{\xi^{2}(T_{\rm obs})},
\end{equation}
where $\xi$ accounts for low-energy free--free photons that are Comptonized to energies of order $kT_{\rm obs}$.

For $\beta_{\rm bo}=0.08$, Eq.~\ref{Eq_eta_th} becomes
\begin{equation}
\eta_{\rm th,bo}
\simeq
5.4
\left(
\frac{kT_{\rm BB, bo}}
{0.12\ {\rm keV}}
\right)^{-1/2}.
\end{equation}
Substituting this expression into the temperature relation above, we obtain $\xi(T_{\rm obs})\simeq5.4$ for
$kT_{\rm obs}=0.12\ {\rm keV}$.
Using the Comptonization correction factor
$\xi=\frac{1}{2}\ln y_{\max}(1.6+\ln y_{\max})$,
where $y_{\max}$ is the maximum Comptonization factor for free--free photons,
we obtain $y_{\max}\simeq13.2$. Together with
$y_{\max}=3(\rho_{\rm ps}/10^{-9}\ {\rm g\ cm^{-3}})^{-1/2}
(kT_{\rm obs}/100\ {\rm eV})^{9/4}$
and $\rho_{\rm ps}=\chi\rho_0$, this yields
$\rho_0\simeq1.7\times10^{-11}\ {\rm g\ cm^{-3}}$. The radiation-dominated shock condition then gives
\begin{equation}
kT_{\rm BB, bo}
\simeq
0.037\ {\rm keV}.
\end{equation}
Substituting this temperature back into the thermal-coupling parameter gives
$\eta_{\rm th,bo}\simeq9.8$, and hence
$T_{\rm obs}/T_{\rm BB, bo}\simeq3.3$, consistent with the result in \citet{Yuan2026}. Although the radiation is out of thermal equilibrium, it may still maintain a Wien-like spectrum, making this solution {consistent with the observed thermal X-ray spectrum}. 

The width of the SBO region follows from Eq.~\ref{Eq_DeltaR_compact}:
\begin{equation}
\Delta R_{\rm bo}
\simeq
3.7\times10^{12}\ {\rm cm},
\qquad
\frac{\Delta R_{\rm bo}}
{R_{\rm bo}}
\simeq
0.18.
\label{delta R_bo/wind}
\end{equation}
Taken together, these results reveal a clear contrast between the two solutions. The local thermal equilibrium case requires a very thin, dense SBO region (note that this does not imply that the full CSM shell has such a small radial width). In contrast, the photon-starved solution requires a broader, lower-density SBO region. Both solutions can reproduce the observed X-ray spectrum and radiated energy. However, the two solutions imply different CSM properties and therefore different predictions for the subsequent  cooling emission, as discussed in Sect.~\ref{sec:csm_cooling}.

{ The  spectra of SN~2026gzf show line velocities of $35000 {\rm \, km \, s^{-1}}$ ($\sim0.1c$) at $\sim4$ days after explosion \citep{OConnor2026}.  The difference between the spectroscopic and the breakout velocities may  reflect the fact that the central-engine activity continued to supply energy to the SN ejecta after the shock breakout. Such energy injection, as suggested by the magnetar-powered light-curve modeling of \citet{Yuan2026}, could accelerate part of the ejecta and yield higher velocities by the epoch of the spectroscopic observations. Thus, the ejecta velocity measured several days later does not need to be identical to the shock velocity at the breakout. }

\subsection{Constraints from the radio data}
\label{subsec:radio_velocity}

Radio observations provide an independent constraint on the shock velocities.
The radio emission traces the kinetic energy of the fastest post-breakout ejecta as they interact with the external medium and drive a synchrotron-emitting forward shock.
At a fixed time, a faster shock has a higher post-shock energy density, a stronger magnetic field, and a larger swept-up mass of radiating electrons.
The radio luminosity is therefore highly sensitive to the velocity of the fastest ejecta expanding into the external medium.

Shock-breakout ejecta are expected to have a steep velocity structure.
Near the stellar surface, the pre-explosion density can be approximated as $\rho_*\propto x^{n}$, where $x=(R_*-r)/R_*$ and $n$ is typically in the range $1-3$.
The cumulative kinetic energy above a given ejecta velocity falls off steeply with increasing velocity.
In the nonrelativistic limit, following \citet{Matzner1999,Tan2001,Duran2015}, we write the cumulative kinetic energy of the ejecta above velocity $\beta c$ as
\begin{equation}
E(>\beta)
\propto
(\beta)^{-s'+1},
\qquad
s'=4.35+\frac{5.35}{n},
\end{equation}
or equivalently $E(>\beta)\propto(\beta)^{-q_{\rm ej}}$, where $q_{\rm ej}=s'-1=3.35+5.35/n$.
For $n=3$, this gives $q_{\rm ej}\simeq5.1$, so even modest deceleration of the fastest material allows slower ejecta, carrying substantially more energy, to refresh the shock (see Appendix~\ref{app:radio_model}).


The radio data show a strong contrast between the two SBO events. 
XRO~080109 is radio bright, with mJy-level detections at tens of GHz during the first $\sim10$ days and subsequent radio coverage  from 0.325 to 95 GHz \citep{Soderberg2008,vanderHorst2011}. 
Its radio spectra and light curves are well described by synchrotron self-absorbed emission from a shock propagating through a stellar wind. 
In contrast, EP260321a does not have a secure radio counterpart. 
Follow-up observations from 0.7 to 23 GHz yielded no secure counterpart over $\sim5$--31 days, with additional VLA monitoring extending to $\sim55$ days. The deepest limits reach $\lesssim20$--$30\,\mu{\rm Jy}$ at 6--22 GHz \citep{Yuan2026,Rastinejad2026}. In terms of spectral luminosity, the contrast is approximately
\begin{equation}
L_\nu \sim
\begin{cases}
10^{27}\mbox{--}10^{28}\ {\rm erg\ s^{-1}\ Hz^{-1}},
& \text{XRO~080109},\\
\lesssim {\rm few}\times10^{26}\ {\rm erg\ s^{-1}\ Hz^{-1}},
& \text{EP260321a}.
\end{cases}
\end{equation}
Thus, despite the fact that its  X-ray breakout energy is larger by a factor of $\sim40$, EP260321a is substantially fainter in the radio than XRO~080109.

Since the synchrotron luminosity depends strongly on the velocity of the fastest ejecta, the weak radio emission of EP260321a favors a slower forward shock than that inferred for XRO~080109. We quantify this with the above refreshed-shock synchrotron framework  (see also Appendix~\ref{app:radio_model}).
The fastest shell with energy $E_{\rm f,0}$ and velocity $\beta_{\rm f,0}$ expands freely at the beginning; after deceleration, slower and more energetic ejecta refresh the forward shock according to Eq.~\ref{Eq_Esh_radio}.
The radio data therefore jointly constrain the wind density, ejecta energy, microphysical parameters, and the velocity of the fastest shell.

\begin{figure*}[htbp]
    \centering
    \includegraphics[width=0.6\linewidth]{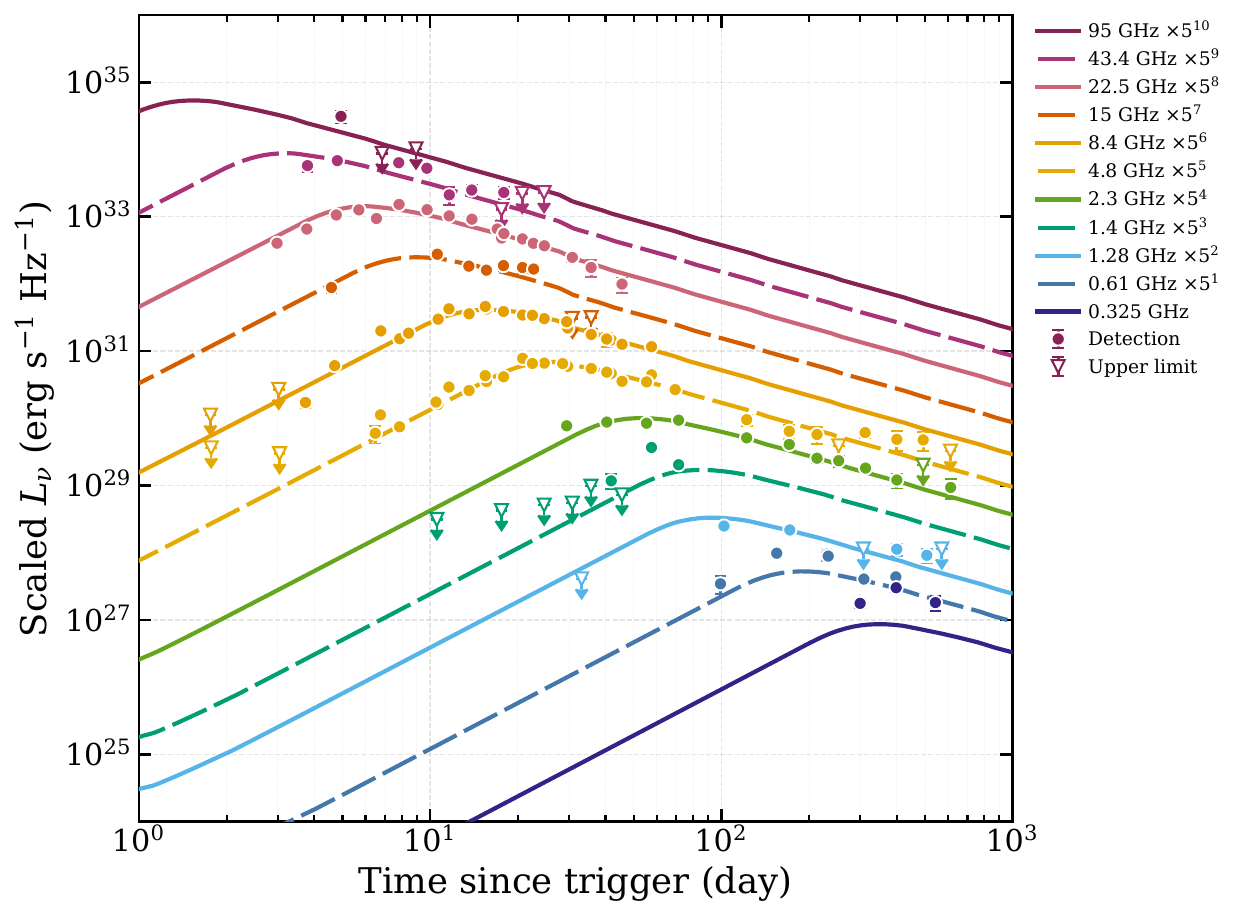}\\
    \includegraphics[width=0.6\linewidth]{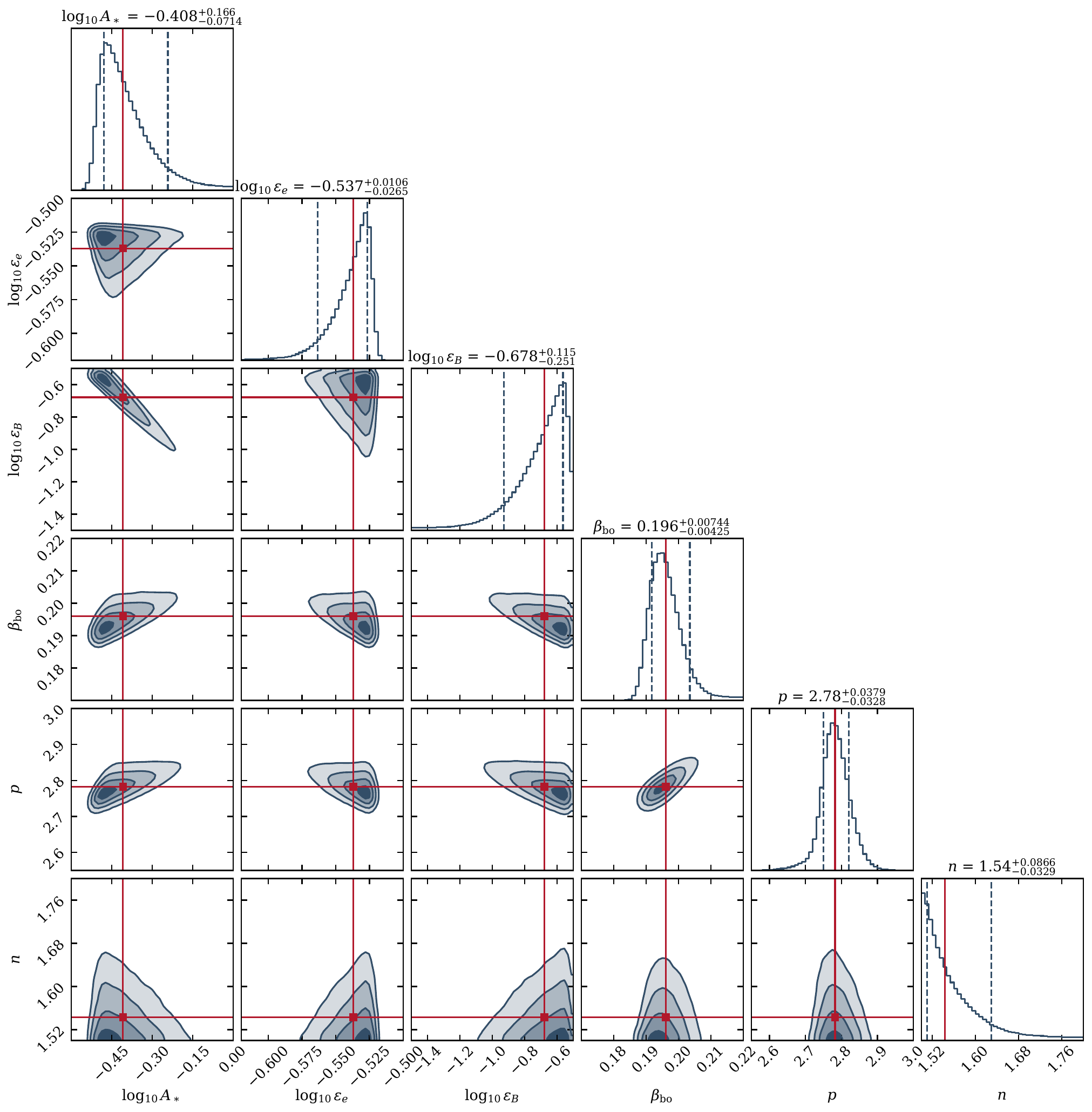}
\caption{\textbf{Modeling of the radio light curves  of XRO~080109/SN~2008D.} Top: multi-frequency spectral-luminosity light curves from 0.325 to 95 GHz. Filled circles indicate detections, downward-pointing triangles denote upper limits, and colored curves show the best-fitting synchrotron models. For clarity, the 0.325 GHz light curve is unscaled, while successive higher-frequency light curves are vertically offset by factors of 5. Bottom: one- and two-dimensional marginalized posterior distributions for the six model parameters, $A_*$, $\epsilon_e$, $\epsilon_B$, $\beta_{\rm bo}$, $p$, and $n$. The quoted values are the posterior medians with 16th--84th percentile uncertainties. Radio data are taken from \citet{Soderberg2008,vanderHorst2011}.}

    \label{fig_XRT080109_modeling}
\end{figure*}

We first use XRO~080109/SN~2008D as a reference ordinary supernova shock-breakout event. We adopt the radio measurements from \citet{Soderberg2008, vanderHorst2011} and include frequency-dependent flux-calibration uncertainties in the error budget. For the VLA/VLBA and CARMA data, we follow the NRAO VLA Observational Status Summary,\footnote{\url{https://science.nrao.edu/facilities/vla/docs/manuals/oss/referencemanual-all-pages}} the NRAO VLBA Observational Status Summary,\footnote{\url{https://science.nrao.edu/facilities/vlba/docs/manuals/oss/amp-cal}}  and the CARMA flux-calibration memo,\footnote{\url{https://pages.astro.umd.edu/~mpound/mmarray.org/memos/carma_memo59.pdf}} respectively. For the WSRT and GMRT measurements, we adopt a 10\% flux-scale uncertainty following \citet{Stroe2016}, who use the same value for WSRT/GMRT radio data. We take the kinetic energy of the shocked wind/CSM at the end of breakout to be $E_{\rm f,0}\sim E_X=2\times10^{46}\ {\rm erg}$. 
In this extended-wind scenario, the internal energy stored in the shocked wind at $R<R_{\rm bo}$ is comparable to that deposited in the wind swept up from the SBO region, $R_{\rm bo}<R\lesssim2R_{\rm bo}$. Since the post-shock internal and kinetic energies are comparable to within a factor of order unity, this motivates $E_{\rm f,0}\sim E_X$.
We fit the multi-frequency radio light curves using a Markov chain Monte Carlo (MCMC) analysis, with six free parameters: the wind-density normalization $A_*$, the post-shock energy fractions in relativistic electrons and magnetic fields, $\epsilon_e$ and $\epsilon_B$, the fastest-shell velocity $\beta_{\rm bo}$, the electron index $p$, and the outer-ejecta structure index $n$. 
Fig.~\ref{fig_XRT080109_modeling} shows the best-fitting model together with the posterior range $16$--$84\%$. A single forward shock propagating through a stellar wind reproduces the multi-frequency radio light curves with synchrotron self-absorbed emission, including the early high-frequency peak and the later low-frequency rise. The fit favors a normalization of the wind-density $A_*=0.39^{+0.18}_{-0.06}$.
{This is well below the  value derived from the breakout emission, $A_{*,{\rm bo}}\simeq c t_{\rm bo}/(5\times10^{11}\kappa)\simeq 24\,t_{\rm bo,80}\kappa_{0.2}^{-1}$, a tension discussed by \citet{Balberg2011} and \citet{Svirski2014}.
It can be reconciled if the shock  breakout is related to the radially compact and enhanced wind, with $R_{\rm w}\lesssim10^{14}\ {\rm cm}$ as suggested by \citet{Svirski2014}.
In this case, the mass in $R_{\rm d}<R<R_{\rm w}$ is smaller than the swept-up mass at the radio-emitting epochs and should have little impact on our  radio afterglow emission calculation.}
The fit also gives $\epsilon_e=0.29^{+0.01}_{-0.02}$, $\epsilon_B=0.21^{+0.06}_{-0.09}$, and an electron index $p=2.78^{+0.04}_{-0.03}$, typical of Ib/c SN shocks.
The inferred outer-ejecta index, $n=1.54^{+0.09}_{-0.03}$, implies a relatively strong increase in kinetic energy toward lower velocities. For a wind-like medium with $k=2$, this corresponds to a refreshed-shock energy evolution $E\propto t^{0.7}$ in the nonrelativistic limit (see Appendix~\ref{app:radio_model}), slightly shallower than  $E\propto t^{0.8}$  inferred for XRO~080109/SN~2008D by \citet{Soderberg2008}. This difference may arise from different choices of the microphysical parameters of $\epsilon_e$ and $\epsilon_B$. Fig.~\ref{fig_XRT080109_modeling} implies a fastest-shell velocity of
\begin{equation}
\beta_{\rm bo}
=
0.20\pm 0.01.
\label{Eq_beta_radio_080109}
\end{equation}
Thus, the radio light curves of XRO~080109/SN~2008D are consistent with a fast but nonrelativistic shock in a wind environment, with late-time evolution shaped by continued energy injection from slower ejecta. 

For comparison, if we fix $\epsilon_e=\epsilon_B=0.1$, as assumed by \citet{Soderberg2008}, the fit gives
$A_*\simeq1.1$ and $\beta_{\rm bo}=0.25 \pm 0.01$, broadly consistent with the above fit that assumes free values of $\epsilon_e$ and $\epsilon_B$.

\begin{figure*}[htbp]
    \centering
    \includegraphics[width=0.8\linewidth]{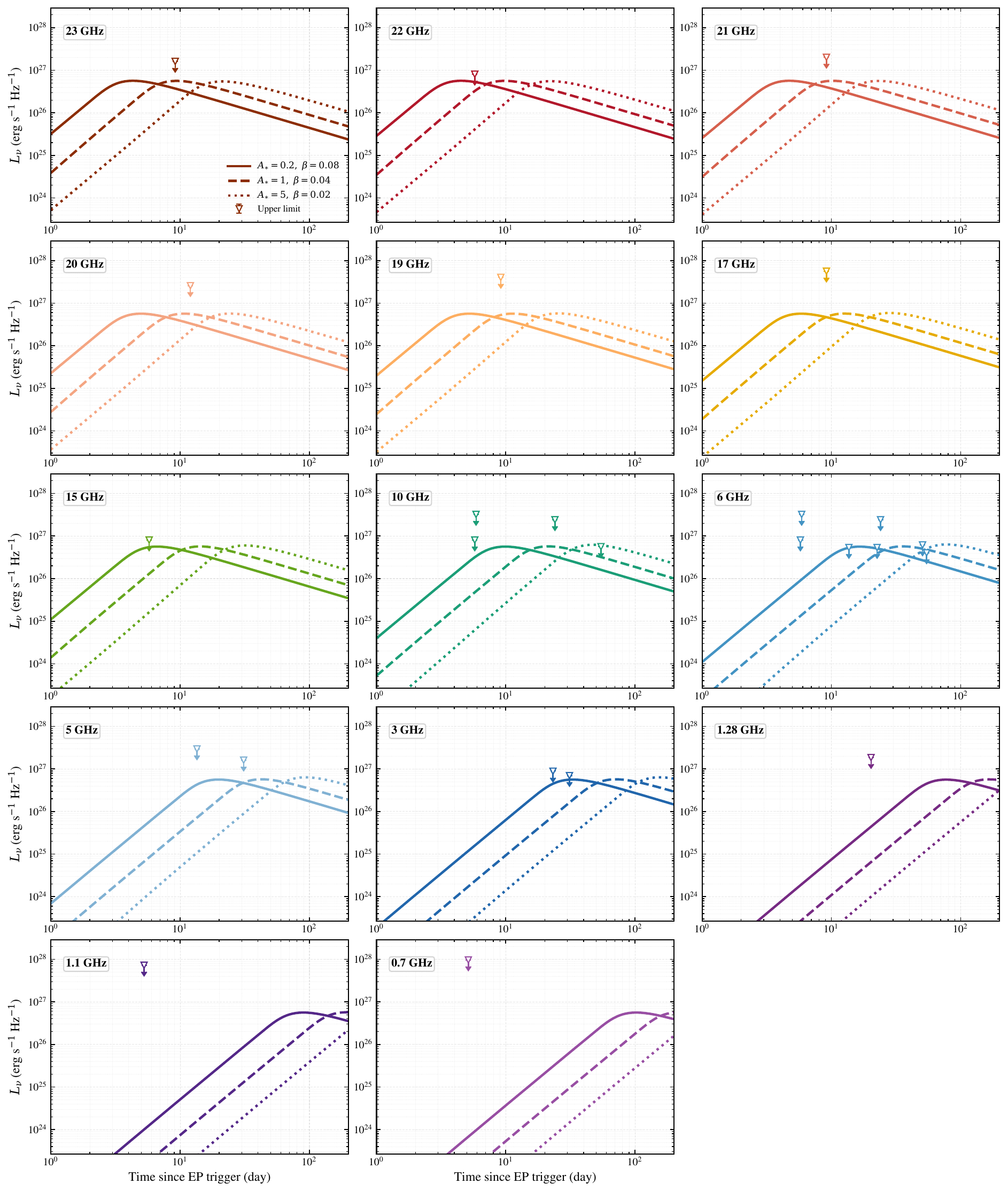}
    \caption{\textbf{Radio constraints on EP260321a.} Multi-frequency radio upper limits are compared with synchrotron emission from a single forward shock propagating in a wind environment. Each panel shows one observing frequency. The three curves correspond to representative combinations of wind density and shock velocity, $A_*=0.2$ with $\beta=0.08$, $A_*=1$ with $\beta=0.04$, and $A_*=5$ with $\beta=0.02$, with the remaining parameters fixed to the best-fitting values. The plotted radio upper limits are taken from \citet{Yuan2026} and \citet{Rastinejad2026}.}
    \label{fig_EP260321a_radio_lc}
\end{figure*}

\begin{figure*}[htbp]
    \centering
    \includegraphics[width=0.6\linewidth]{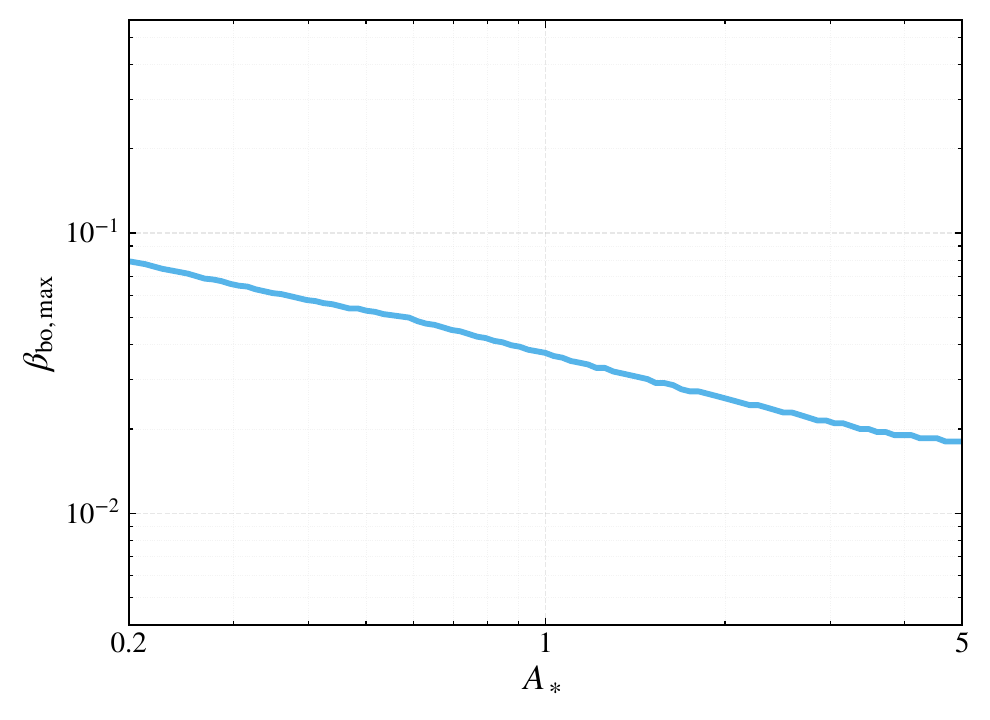}
    \caption{\textbf{Radio constraint on the fastest-shell velocity of EP260321a.} Maximum allowed velocity $\beta_{\rm bo,max}$ as a function of the wind-density normalization $A_*$. For each $A_*$, $\beta_{\rm bo,max}$ is obtained by requiring the synchrotron emission from a single forward shock propagating in a wind environment not to exceed any radio upper limit. The remaining parameters are fixed to the best-fitting values inferred from XRO~080109/SN~2008D, adopted here as a reference ordinary supernova shock-breakout event.}
    \label{fig_EP260321a_radio_beta}
\end{figure*}

Unlike the extended-wind case, the compact-shell geometry can allow a much larger reservoir of trapped energy if
$\Delta R_{\rm CSM}\gg\Delta R_{\rm bo}$,
as inferred below from the optical cooling emission of EP260321a. The observed X-rays trace only the energy deposited in the material swept up from the thin SBO region, $E_X\sim4\pi R_{\rm bo}^{2}\rho_0\Delta R_{\rm bo}v_{\rm bo}^{2}$, whereas the internal energy of the entire shocked CSM shell, $E_{\rm int,0}\sim4\pi R_{\rm bo}^{2}\rho_0\Delta R_{\rm CSM}v_{\rm bo}^{2}$, remains trapped and later powers the cooling emission (Fig.~\ref{fig_shock_breakout_schematic}). Thus, $E_X$ provides only a lower limit on the shocked-CSM energy. Nevertheless, for EP260321a, adopting a higher value of $E_{\rm f,0}$ has little effect on the radio constraint because the fastest ejecta remain in the coasting phase during the radio observation epoch. To allow for a different wind density in EP260321a, we vary the  normalization $A_*$ while fixing the remaining parameters ($\epsilon_e$, $\epsilon_B$, $p$, and $n$) to the best-fit values inferred from XRO~080109/SN~2008D. 
For each $A_*$, we determine the maximum fastest-shell velocity, $\beta_{\rm bo,max}$, for which the predicted radio emission remains below all  upper limits. 
The allowed velocity decreases monotonically with wind density; over the range $A_*\simeq0.2$--$5$, we find
\begin{equation}
\beta_{\rm bo}\lesssim 0.02\mbox{--}0.08,
\label{Eq_beta_radio_260321}
\end{equation}
as shown by the radio light-curve comparison in Fig.~\ref{fig_EP260321a_radio_lc} and the $\beta_{\rm bo,max}$--$A_*$ dependence in Fig.~\ref{fig_EP260321a_radio_beta}. {For comparison, if we adopt fixed microphysical parameters $\epsilon_e=\epsilon_B=0.1$, as in \citet{Soderberg2008}, we find $\beta_{\rm bo}\lesssim0.03$--$0.13$ over the same range of $A_*$, which is  below the corresponding value inferred for XRO~080109/SN~2008D, $\beta_{\rm bo} = 0.25 \pm 0.01$.} Thus,  radio nondetections require a lower velocity for the fastest shell in EP260321a than that in XRO~080109/SN~2008D, consistent with the slow shock inferred independently from the X-ray emission. This velocity contrast also naturally explains the different X-ray spectra: EP260321a shows a thermal blackbody-like spectrum, whereas XRO~080109 is dominated by nonthermal emission, as expected if efficient thermalization occurs only for relatively slow shocks, $\beta_{\rm bo}\lesssim0.1$ \citep{Katz2010,Nakar2010}.

\section{The Mass of  CSM and the Cooling Emission}
\label{sec:csm_cooling}

In addition to the shock velocity, the CSM mass is another key distinction between EP260321a and XRO~080109. A massive CSM located at a large radius naturally produces bright cooling emission from the shocked CSM, which can be reflected in the early optical emission before $\sim 1$ day. 

For XRO~080109/SN~2008D, the early ultraviolet/optical emission showed two components. 
The early blue component peaked before one day after the X-ray outburst and was modeled as cooling emission from the shock-heated envelope of a compact Wolf--Rayet (WR) progenitor \citep{Soderberg2008,Chevalier2008,Modjaz2009,Rabinak2011}. The later redder component peaked at $\sim20$ days and was powered by radioactive decay \citep{Soderberg2008}.

EP260321a/SN~2026gzf exhibits markedly different early optical evolution. Its optical emission remains bright and nearly flat over $t\simeq0.1$--$1$ day, with $L_\nu\sim\text{a few}\times10^{27}\ {\rm erg\ s^{-1}\ Hz^{-1}}$, about a factor of $\sim 10$ higher than that of XRO~080109/SN~2008D at $t\sim0.1$ day and  a factor of $\sim 3$ higher at $t\sim1$ day \citep{Soderberg2008}. The later, broader SN component rises after $\sim1$ day and peaks at $\sim10$--$15$ days, consistent with emission from SN ejecta powered mainly by radioactive decay, with a possible contribution from the CSM interaction and a newborn magnetar \citep{Yuan2026,Chen2026,Rastinejad2026,Martin-Carrillo2026}. The early optical component of EP260321a is too bright to be cooling emission from a bare compact Wolf--Rayet core. Instead, it points to cooling emission from extended shock-heated material around the progenitor, as expected for a compact core surrounded by a dense CSM \citep{NakarPiro2014,Piro2015}. 
After crossing the stellar core, the radiation-mediated shock propagates into the extended CSM and heats it. 
{Because the CSM is shock-heated at a radius larger than that of a compact WR envelope, the trapped radiation suffers only modest adiabatic losses before escaping, allowing the cooling emission to dominate the early light curve.} This emission drops rapidly once the entire shocked CSM becomes optically thin, so for low CSM  masses the cooling phase typically lasts only hours to days \citep{NakarPiro2014}.

The difference in the amount of CSM can be estimated directly from the SBO emission. 
For XRO~080109, the extended-wind solution gives
\begin{equation}
    A
    =
    \frac{c t_{\rm bo}}{\kappa}
    =
    1.2\times10^{13}\,
    t_{\rm bo,80}\,
    \kappa_{0.2}^{-1}
    \ {\rm g\ cm^{-1}}.
\end{equation}
The wind mass involved in the shock breakout is then 
\begin{equation}
\label{csm-mass-080109}
    M_{\rm w,bo}
    \sim
    4\pi A R_{\rm bo}
    =
    5.6\times10^{-8}\,
    \beta_{\rm bo,0.3}\,
    t_{\rm bo,80}^{2}\,
    \kappa_{0.2}^{-1}
    \ M_\odot .
\end{equation}
Such a small CSM mass is insufficient to produce an observable day-scale CSM-cooling component. The relatively weak early optical emission of XRO~080109/SN~2008D is instead attributed to cooling emission of the shock-heated stellar envelope \citep{Soderberg2008, Chevalier2008, Modjaz2009,Rabinak2011}.

For EP260321a, the X-ray breakout properties allow two interpretations: a slow shock in local thermal equilibrium and a faster, photon-starved shock that is out of thermal equilibrium (Sect.~\ref{subsec:xray_velocity}). We consider both cases below because they predict different levels of optical cooling emission.

\paragraph{\bf The local thermal equilibrium case.}
For the  thermal equilibrium case, we need
$\beta_{\rm bo}\lesssim0.05$. The corresponding breakout properties, including
$\rho_0$, $R_{\rm bo}$, and $\Delta R_{\rm bo}$, are derived in Sect.~\ref{subsec:xray_velocity}. The mass contained in the thin SBO region is
\begin{align}
m_{\rm bo}
&=
4\pi R_{\rm bo}^{2}\rho_0\Delta R_{\rm bo}
\notag\\
&\simeq
4.5\times10^{-4}
E_{\rm bo,48}
\beta_{\rm bo,0.05}^{-2}
M_\odot .
\label{Eq_m_bo_260321}
\end{align}
Here we have used Eq.~\ref{Eq_beta_bo_260321}, taking $f_{\rm sh}=1$, to express $R_{\rm bo}$ in terms of $E_{\rm bo}$ and $\beta_{\rm bo}$.

For a uniform shell, the total CSM mass depends on the radial width $\Delta R_{\rm CSM}$ of the shell,
\begin{align}
M_{\rm CSM}
&=
4\pi\rho_0R_{\rm bo}^{3}f_\Delta
\notag\\
&\simeq
0.06
\kappa_{0.2}^{3/2}
E_{\rm bo,48}^{3/2}
\beta_{\rm bo,0.05}^{-7/2}
f_{\Delta,-1}
\left(
\frac{kT_{\rm BB, bo}}
{0.12\,{\rm keV}}
\right)^{4}
M_\odot ,
\label{Eq_M_CSM_260321}
\end{align}
where $f_\Delta\equiv\Delta R_{\rm CSM}/R_{\rm bo}$ and
$f_{\Delta,-1}\equiv f_\Delta/0.1$.
Thus, even for a moderate fractional shell width, the total CSM mass can be much larger than the mass contained in the SBO region. {We note that the density profile interior to the dense CSM shell has a negligible effect on our calculation of the optical cooling emission, provided that the total mass in this region is much smaller than $M_{\rm CSM}$. As discussed in more detail in Sect.~\ref{sec:summary}, assuming that the interior region of the dense CSM shell is filled by a stellar wind, the minimum wind mass required to keep the radiation trapped until the shock approaches the shell is only $\sim 10^{-4}\,M_\odot$. 
This is much smaller than $M_{\rm CSM}\simeq0.06\,M_\odot$, so the inner wind would make a negligible contribution to the subsequent cooling emission.
}

The initial diffusion and expansion times of the shocked CSM at breakout are, respectively (see Appendix~\ref{app:cooling_model}),
\begin{align}
t_{d,0}
&=
\frac{\kappa\rho_0R_{\rm bo}^{2}f_\Delta^{2}}
{\chi c}
\notag\\
&\simeq
0.38\,
\kappa_{0.2}^{2}
E_{\rm bo,48}
\beta_{\rm bo,0.05}^{-3}
f_{\Delta,-1}^{2}
\left(
\frac{kT_{\rm BB, bo}}
{0.12\,{\rm keV}}
\right)^{4}
\ {\rm d},
\label{Eq_td0_cooling_scaling}
\\
t_{\rm exp}
&=
\frac{R_{\rm bo}}{v_{\rm bo}}
\notag\\
&\simeq
0.21\,
\kappa_{0.2}^{1/2}
E_{\rm bo,48}^{1/2}
\beta_{\rm bo,0.05}^{-3/2}
\ {\rm d}.
\label{Eq_texp_cooling_scaling}
\end{align}

As the shocked CSM expands, its optical depth decreases and the diffusion time becomes shorter. The trapped radiation is released efficiently when this diffusion time becomes comparable to the time since breakout. We therefore define a characteristic bolometric release time $t_{\rm p}$ by
$t_{\rm diff}(t)\simeq t$ (see Appendix~\ref{app:cooling_model}):
\begin{align}
t_{\rm p}
&=
\frac{
\left(t_{\rm exp}^{2}+4t_{d,0}t_{\rm exp}\right)^{1/2}-t_{\rm exp}
}{2}
\notag\\
&=
\Xi(q)\,t_{d,0}
\notag\\
&\simeq
0.38\,
\Xi(q)\,
\kappa_{0.2}^{2}
E_{\rm bo,48}
\beta_{\rm bo,0.05}^{-3}
f_{\Delta,-1}^{2}
\left(
\frac{kT_{\rm BB, bo}}
{0.12\,{\rm keV}}
\right)^{4}
\ {\rm d},
\label{Eq_tp_cooling_scaling}
\end{align}
where
$q\equiv t_{d,0}/t_{\rm exp}$ and
$\Xi(q)\equiv[(1+4q)^{1/2}-1]/(2q)$.
For the fiducial parameters, $q\simeq1.9$ and
$\Xi(q)\simeq0.51$, we obtain
$t_{\rm p}\simeq0.20\ {\rm d}$.

Accounting for adiabatic losses before photon escape, the bolometric luminosity at the characteristic release time $t_{\rm p}$ is
(Appendix~\ref{app:cooling_model})
\begin{align}
L_{\rm p}
&=
\frac{E_{\rm int,0}}{t_{d,0}}
\exp\left[
-\Xi(q)-\frac{q\Xi^{2}(q)}{2}
\right]
\notag\\
&\simeq
4.0\times10^{45}\,
\kappa_{0.2}^{-1/2}
E_{\rm bo,48}^{1/2}
\beta_{\rm bo,0.05}^{3/2}
f_{\Delta,-1}^{-1}
\notag\\
&\times \exp \left[
-\Xi(q)-\frac{q\Xi^{2}(q)}{2}
\right]
\ {\rm erg\ s^{-1}} .
\label{Eq_Lp_cooling_scaling}
\end{align}
Here $E_{\rm int,0}\simeq M_{\rm CSM}v_{\rm bo}^{2}/2$ is the initial internal energy of the shocked CSM. For the fiducial parameters, $q\simeq1.9$ and
$\Xi(q)\simeq0.51$, the exponential factor is $\simeq0.47$, and we obtain
$L_{\rm p}\simeq1.9\times10^{45}\ {\rm erg\ s^{-1}}$.

The optical-band maximum occurs later than the bolometric peak. As the shocked CSM cools, the bolometric luminosity declines, while the spectral peak shifts into the optical band, causing the optical luminosity to continue rising. The optical peak is reached when these two effects are balanced. Quantitatively, solving $d\ln L_\nu/dt=0$ in the Rayleigh--Jeans limit gives (Appendix~\ref{app:cooling_model})
\begin{align}
t_{\nu,p}
&=
t_{\rm exp}
\left[
\left(
\frac{6t_{d,0}}{t_{\rm exp}}
\right)^{1/2}
-1
\right]
\notag\\
&\simeq
\left(
6t_{d,0}t_{\rm exp}
\right)^{1/2}
\notag\\
&\simeq
0.7\,
\kappa_{0.2}^{5/4}
E_{\rm bo,48}^{3/4}
\beta_{\rm bo,0.05}^{-9/4}
f_{\Delta,-1}
\left(
\frac{kT_{\rm BB, bo}}
{0.12\,{\rm keV}}
\right)^{2}
\ {\rm d}.
\label{Eq_tnup_cooling_scaling}
\end{align}
In the second step, the second term in the brackets has been neglected to obtain a simple analytic scaling.

The corresponding Rayleigh--Jeans peak spectral luminosity is
\begin{align}
L_{\nu,p}^{\rm RJ}
&=
\frac{8\pi^2k_{\rm B}\nu^2}{c^2} \notag\\
&\quad\times 
R_{\rm bo}^{3/2}
\left(
\frac{E_{\rm int,0}}
{4\pi\sigma_{\rm SB}t_{d,0}}
\right)^{1/4}
\Lambda_{\nu,p}^{3/2}
\exp\left[
-\frac{\zeta(\Lambda_{\nu,p}^{2}-1)}{4}
\right]
\notag\\
&\simeq
2.8\times10^{27}
\nu_{14.78}^{2}
\kappa_{0.2}^{7/4}
E_{\rm bo,48}^{5/4}
\notag\\
&\quad\times
\beta_{\rm bo,0.05}^{-3/2}
f_{\Delta,-1}^{5/4}
\left(
\frac{kT_{\rm BB,bo}}
{0.12\,{\rm keV}}
\right)^{3}
\ {\rm erg\ s^{-1}\ Hz^{-1}} .
\label{Eq_Lnup_cooling_scaling}
\end{align}
where
$\nu_{14.78}\equiv\nu/(6\times10^{14}\ {\rm Hz})$,
$\Lambda_{\nu,p}=(6t_{d,0}/t_{\rm exp})^{1/2}$,
and
$\zeta=t_{\rm exp}/(2t_{d,0})$.
For $\beta=0.05$, $E_{\rm bo}=8.7\times10^{47}{\rm erg}$ and $\kappa=0.2$, the approximate scaling gives
$t_{\nu,p}\simeq0.7 f_{\Delta,-1} {\rm d}$.
The corresponding peak luminosity is
$L_{\nu,p}^{\rm RJ}\simeq2.8\times10^{27}\nu_{14.78}^{2}  f_{\Delta,-1}^{5/4}\ {\rm erg\ s^{-1}\ Hz^{-1}}$.
Compared with the observed sub-day optical data of EP260321a, we infer
$f_\Delta\simeq0.1$, corresponding to
$M_{\rm CSM}\simeq0.06M_\odot$. 

\begin{figure*}[htbp]
    \centering
    \includegraphics[width=1.0\linewidth]{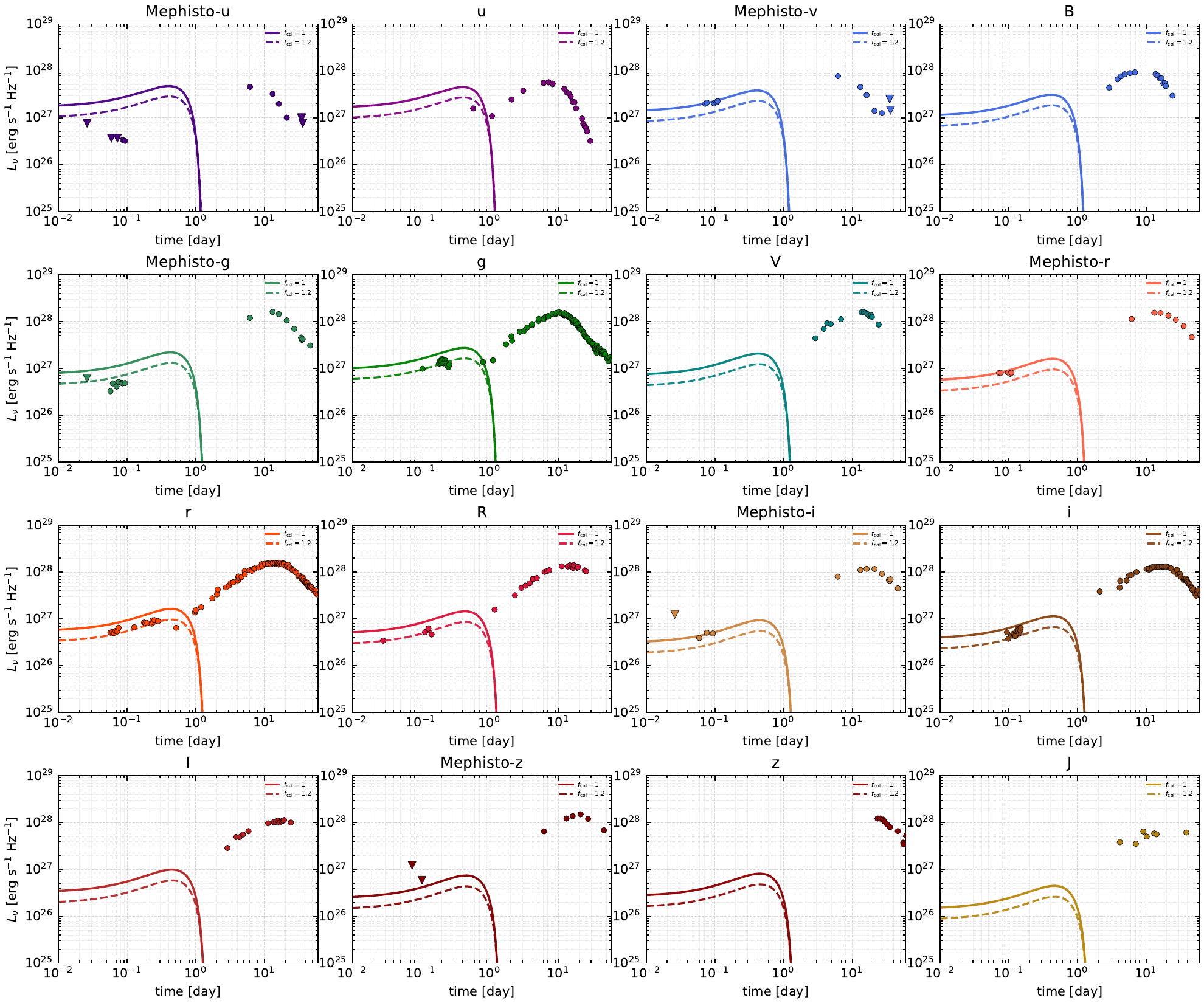}
    \caption{\textbf{Cooling emission from the shocked CSM in EP260321a.} The multiband optical light curves are compared with the thermal emission from trapped radiation diffusing out of the shocked CSM shell after breakout, shown without color correction ($f_{\rm col}=1$; solid line) and with a modest color correction ($f_{\rm col}=1.2$; dashed line). The cooling component can account for the earliest optical emission but fades rapidly, whereas the later rise is associated with SN~2026gzf and requires additional power from the supernova ejecta and/or CSM interaction \citep{Yuan2026,OConnor2026,Martin-Carrillo2026,Chen2026,Rastinejad2026}. Filled circles mark detections, and downward triangles denote upper limits. The plotted photometry is taken from \citet{Yuan2026}, \citet{Chen2026}, and \citet{Martin-Carrillo2026}.}
    \label{fig_EP260321a_cooling}
\end{figure*}

We further calculate the light curve of the cooling emission from the shocked CSM using the numerical method in Appendix~\ref{app:cooling_model}, taking
$f_\Delta=0.1$ and the breakout parameter values derived in Sect.~\ref{subsec:xray_velocity} ($E_{\rm bo}=8.7\times10^{47}\ {\rm erg}$,
$kT_{\rm BB, bo}=0.124\ {\rm keV}$, and
$\beta_{\rm bo}=0.05$).
The resulting multiband light curves are shown in Fig.~\ref{fig_EP260321a_cooling}. As shown in the figure, the shocked-CSM cooling emission can roughly account for the  optical emission of EP260321a at
$t\simeq0.1$--$1$ day. 
{At $t\lesssim0.1$ day, however, the simple blackbody model overpredicts the Mephisto-$u$-band flux by a factor of $\sim3$. This tension arises because, in our cooling-emission model, the effective temperature at $t\simeq0.1$ day remains high, $T_{\rm eff}\sim 2 \times10^5\ {\rm K}$. At such a high temperature, the Mephisto-$v$ band and the redder optical bands lie on the Rayleigh--Jeans tail of the model spectrum, which is broadly consistent with their observed fluxes. The observed Mephisto-$u$ flux, however, falls below the Rayleigh--Jeans extrapolation from these bands. A lower temperature of $T\simeq1.8\times10^4\ {\rm K}$, as assumed in \citet{Yuan2026}, can reduce the Mephisto-$u$-band flux, but cannot simultaneously account for the bright Mephisto-$v$ emission. This suggests that the earliest Mephisto-$u$ and $v$ fluxes cannot be described by a single, unattenuated blackbody spectrum. 
Reproducing the Mephisto-$u$ flux would require a lower temperature, whereas matching the bright Mephisto-$v$ flux would require a higher temperature. We favor the higher-temperature interpretation because it is consistent with our shock-cooling model. The discrepancy between the observed  Mephisto-$u$ band flux and our model prediction  may reflect the limitations of the blackbody approximation, which neglects frequency-dependent opacity in the expanding SN/CSM material \citep[e.g.][]{Dessart2015}. A quantitative treatment of this effect requires multi-frequency radiative-transfer calculations and is beyond the scope of this work.}


\paragraph{\bf Photon-starved spectrum case.}
For comparison, the photon-starved spectrum case corresponds to a high shock breakout velocity, e.g., $\beta_{\rm bo}=0.08$. 
It requires a much lower  density at the breakout radius,
$\rho_0\simeq1.7\times10^{-11}\ {\rm g\ cm^{-3}}$,
and a wider SBO region,
$\Delta R_{\rm bo}\simeq0.18R_{\rm bo}$ (see Eq.~\ref{delta R_bo/wind}). Even if this low-density region extends over a radial width comparable to the width of the SBO region, the corresponding mass is only a few $\times10^{-4}M_\odot$, and the diffusion time through the shocked CSM is
$\sim2\times10^{2}\ {\rm s}$. This mass is too small to produce the observed sub-day optical emission.

Reproducing the bright optical emission in this case would require the CSM inside the SBO region to be much denser than the SBO region itself, with the density increasing inward more steeply than $\rho\propto r^{-2}$. Such a stratified shell may still be possible, but it requires a detailed study, which is beyond the scope of this work. {Alternatively, the early optical emission may be produced by a subrelativistic cocoon \citep{Yuan2026}. }

{We also note that the inferred breakout and cooling properties are  insensitive to the assumed CSM density profile and to whether the ionized CSM is He- or C/O-dominated. 
For XRO~080109, generalizing the extended-wind profile to $\rho\propto r^{-s}$ ($s>1$) and using the same breakout condition  (see Sect.~\ref{subsec:xray_velocity}) gives $v_{\rm bo}\propto(s-1)^{-2/3}$ and $R_{\rm bo}\propto(s-1)^{1/3}$. Profiles with $s$ close to 2 therefore still imply a fast breakout shock. For EP260321a, the inferred shell is sufficiently thin ($f_\Delta\simeq0.1$) that an $r^{-2}$ density profile varies only by $\sim20\%$ across it, making the uniform-density approximation adequate. In addition, the CSM around SN~2008D is likely He-rich, with possible residual hydrogen, while the more strongly stripped progenitor of SN~2026gzf could have produced  He-rich or C/O-rich material. In ionized material, He and C/O provide similar numbers of free electrons per baryon and therefore have nearly identical electron-scattering opacities, consistent with the value $\kappa=0.2\ {\rm cm^2\,g^{-1}}$ adopted here.}

\section{A Unified Interpretation of EP260321a and XRO~080109}
\label{sec:unified_interpretation}

A supernova explosion drives a radiation-dominated shock that propagates through the stellar core and accelerates at the sharply decreasing density profile near the stellar edge, bringing progressively smaller amounts of mass to higher velocities and producing a velocity profile $v(m_{\rm c})$, where $m_{\rm c}$ is the core mass accelerated to velocity $v$. As the shock enters the shallower density profile of the extended CSM, it decelerates and forms a forward--reverse shock structure. During the deceleration, the swept-up CSM mass is comparable to the core mass crossed by the reverse shock, so after the  CSM is shocked, its velocity is $v_{\rm CSM}\simeq v(m_{\rm c}=M_{\rm CSM})$. 
For a  density profile $\rho_*\propto x^{n}$ near the stellar edge, the  self-similar solution gives $v\propto m_{\rm c}^{-\alpha_n}$, with $\alpha_n=0.19 n/(n+1)$ \citep{Sakurai1960,Nakar2010}. 
We therefore write $v(m_{\rm c})=v_{\rm CSM}(m_{\rm c}/M_{\rm CSM})^{-\alpha_n}$. The normalization can be obtained by requiring that the integrated kinetic energy in the core mass be equal to the explosion energy, $E_{\rm SN}\simeq (1/2)\int_0^{M_{\rm c}}v^2(m_{\rm c})\,dm_{\rm c}$. 
Since $\alpha_n<1/2$, this gives
\begin{align}
v_{\rm CSM}
&\simeq
\left[
\frac{2(1-2\alpha_n)E_{\rm SN}}{M_{\rm c}}
\right]^{1/2}
\left(
\frac{M_{\rm CSM}}{M_{\rm c}}
\right)^{-\alpha_n} \notag\\
&\simeq
10^{9}
C_{\alpha,1}
{v_{\rm bulk,9}}
\left(\frac{M_{\rm c}}{M_\odot}\right)^{\alpha_n}
\left(\frac{M_{\rm CSM}}{0.01M_\odot}\right)^{-\alpha_n}
{\rm cm\ s^{-1}},
\label{Eq_Piro_ve_general}
\end{align}
where $M_{\rm c}$ is the  mass of the whole stellar core, $v_{\rm bulk}\equiv(2E_{\rm SN}/M_{\rm c})^{1/2}$, $C_{\alpha,1}\equiv (1-2\alpha_n)^{1/2}10^{2\alpha_n}$, and $v_{\rm bulk,9}=v_{\rm bulk}/(10^9 {\rm cm s^{-1}})$.

The energy deposited into the shocked CSM is then
\begin{align}
E_{\rm CSM}
&\simeq
\frac{1}{2}M_{\rm CSM}v_{\rm CSM}^{2} \notag\\
&\simeq
10^{49}
C_{\alpha,2}
v_{\rm bulk, 9}^2
\left(\frac{M_{\rm c}}{M_\odot}\right)^{2\alpha_n}
\left(\frac{M_{\rm CSM}}{0.01M_\odot}\right)^{1-2\alpha_n}
{\rm erg},
\label{Eq_Piro_Ee_general}
\end{align}
where $C_{\alpha,2}\equiv (1-2\alpha_n)10^{4\alpha_n}$. 


{These scalings indicate that the two explosions may have comparable bulk ejecta velocities}: For the estimates below, we adopt $n\simeq1.5$, as inferred from the radio afterglow modeling of XRO~080109/SN~2008D in Sect.~\ref{subsec:radio_velocity}. This gives $\alpha_n=0.114$, $C_{\alpha,1}=1.49$, and $C_{\alpha,2}=2.21$. Inverting Eq.~\ref{Eq_Piro_ve_general} gives
\begin{equation}
    v_{\rm bulk}
    \simeq
    \frac{10^{9}}{C_{\alpha, 1}}
    v_{\rm CSM, 9}
    \left(\frac{M_{\rm CSM}}{0.01M_\odot}\right)^{\alpha_n}
    \left(\frac{M_{\rm c}}{M_\odot}\right)^{-\alpha_n}
    {\rm cm\ s^{-1}} .
    \label{Eq_vbulk_inferred}
\end{equation}
For EP260321a, using $v_{\rm CSM}\simeq0.05c$ and $M_{\rm CSM}\simeq0.06M_\odot$, we obtain
\begin{equation}
    v_{\rm bulk}
    \simeq
    10^{9}
    \left(\frac{M_{\rm c}}{5M_\odot}\right)^{-0.114}
    {\rm cm\ s^{-1}} .
\end{equation}
For XRO~080109, using $v_{\rm CSM}\simeq0.3c$ and $M_{\rm CSM}\simeq5.6\times10^{-8}M_\odot$, we obtain 
\begin{equation}
    v_{\rm bulk}
    \simeq
    1.3\times10^{9}
    \left(\frac{M_{\rm c}}{5M_\odot}\right)^{-0.114}
    {\rm cm\ s^{-1}} .
\end{equation}
The inferred bulk velocities are therefore comparable to each other within a factor of two, motivating the use of a common fiducial explosion scale for the two events.
{We therefore take $E_{\rm SN}=3\times10^{51}\,{\rm erg}$
and $M_{\rm c}=5M_\odot$ as the fiducial parameters for both events. Adopting $M_{\rm CSM}=0.06M_\odot$ for EP260321a and $M_{\rm CSM} \sim 10^{-7}M_\odot$ for XRO~080109, as inferred in Sect.~\ref{sec:csm_cooling}, Eqs.~\ref{Eq_Piro_ve_general} and \ref{Eq_Piro_Ee_general} give}
\begin{align}
    v_{\rm CSM}
    &\simeq
    \begin{cases}
    1.1\times10^{9}\ {\rm cm\ s^{-1}}
    \simeq 0.04c, & {\rm EP260321a},\\
    5.1\times10^{9}\ {\rm cm\ s^{-1}}
    \simeq 0.2c, & {\rm XRO~080109},
    \end{cases}
    \\
    E_{\rm CSM}
    &\simeq
    \begin{cases}
    10^{50}\ {\rm erg}, & {\rm EP260321a},\\
    {\rm few}\times10^{45}\ {\rm erg}, & {\rm XRO~080109}.
    \end{cases}
\end{align}
For EP260321a, the velocity is close to that inferred from the thermal X-ray emission and consistent with the constraint by the radio non-detections. The deposited energy is also comparable to that required by the cooling-emission calculation in Sect.~\ref{sec:csm_cooling}.

For XRO~080109, the velocity is consistent with that inferred from the radio emission and X-ray emission, but the available CSM-cooling energy is quite low. The early optical emission of XRO~080109 should therefore be produced by other processes, such as the cooling emission of the shock-heated stellar envelope of the WR star \citep{Soderberg2008,Chevalier2008,Modjaz2009,Rabinak2011}.

The key difference between the two SBO events therefore lies in their final-stage mass-loss environments, with EP260321a an SBO from a much more massive CSM. The larger CSM mass decelerates the outermost SN ejecta to a lower  velocity and stores more shock-heated internal energy, which  later emerges as the cooling emission, powering the early optical emission in EP260321a.

{ We note that $ v_{\rm CSM}$ here refers to the shock velocity at CSM breakout and is distinct from that at the stellar surface. Although the compact C/O progenitor of SN~2026gzf may have produced a faster shock near the stellar surface than the He-bearing progenitor of SN~2008D, the shock is subsequently decelerated by the massive CSM shell before it breaks out. }

\section{Summary and Discussion}
\label{sec:summary}

We conducted a comparative study of EP260321a and XRO~080109, which are believed to be SBOs from stripped-envelope supernovae. Although both can be interpreted as regular SN shock breakouts, their observed properties differ markedly. EP260321a has higher energy in X-rays, a softer thermal spectrum,  weaker radio emission, and  brighter optical emission within the first day. We have shown that these differences are driven mainly by the shock velocity and the mass of CSM surrounding the progenitor.

The shock velocity is independently constrained by the X-ray breakout emission and the radio data. 
For XRO~080109, the duration and energy of the X-rays imply shock breakout from the surrounding wind with $\beta_{\rm bo}\sim0.4$, subject to a factor of two uncertainty in the breakout radius. The radio light curves are reproduced assuming a  velocity of $\beta_{\rm bo}\simeq0.2$ for the fastest ejecta, broadly consistent with the X-ray-based estimate. 
{For EP260321a, the X-ray outburst and early optical emission suggest an SBO event emerging from the outermost edge of a compact, dense CSM shell, with a lower shock velocity of $\beta_{\rm bo}\simeq0.05$. The radio non-detections independently favor a low velocity for the fastest ejecta, with the allowed upper limit depending on the wind density and microphysical parameters.
The earliest spectra of SN~2026gzf show line velocities of $\sim0.1c$ at $\sim4$ days after explosion \citep{OConnor2026}. 
The difference between the spectroscopic and breakout velocities may  reflect continued central-engine activity after the shock breakout. Such energy injection, as suggested by the magnetar-powered light-curve modeling of \citet{Yuan2026}, could accelerate part of the ejecta and yield higher velocities by the epoch of the spectroscopic observations.} 


The second key difference is the CSM mass. For XRO~080109, {the wind mass involved in the shock breakout can be directly derived from the SBO energy, resulting in  $\sim10^{-7}M_\odot$.} By contrast, EP260321a requires a much more massive CSM shell, as inferred from its early optical cooling emission. For EP260321a, the observed X-rays trace only the energy deposited in the material of the geometrically thin SBO region, whereas most of the shocked-CSM energy remains trapped. From the early optical emission, we infer a CSM  mass of $M_{\rm CSM}\simeq0.06M_\odot$. The energy in the shocked CSM is $E_{\rm int,0} \sim 10^{50} \,  {\rm erg}$, which powers the bright optical emission at $t\simeq0.1$--$1$ day when the radiation in the shocked CSM can diffuse out. This trapped energy reservoir does not affect the radio constraints, because the fastest ejecta are still in the coasting phase during the radio observation epochs.


The lower shock velocity and the larger CSM mass in EP260321a could be
unified in a picture in which  a larger CSM mass decelerates the radiation-mediated shock to a lower velocity before it breaks out.
The differences between these two events may therefore reflect the final-stage mass-loss behavior rather than a  difference in the underlying SN explosion. The low CSM mass in XRO~080109 may be due to an enhanced but steady wind loss before  shock breakout, while the high CSM mass  in EP260321a could be due to sudden ejection of the outer envelope shells prior to the SN explosion.  



{{ In our fiducial calculation, material interior to the dense CSM shell has a negligible effect on the cooling emission from the shell, provided that its total mass is much smaller than $M_{\rm CSM}$.} If this region is instead filled by a stellar wind, the wind can sustain the radiation-mediated shock as long as its optical depth satisfies $\tau_{\rm w}\gtrsim c/v_{\rm sh}$. For a steady wind with $\rho=A r^{-2}$, the minimum wind mass required to keep the radiation trapped until the shock approaches the dense CSM shell is $M_{\rm w,min}(<R_{\rm bo})\simeq4\pi R_{\rm bo}^{2}/(\kappa\beta_{\rm sh})\lesssim4\pi R_{\rm bo}^{2}/(\kappa\beta_{\rm bo})\simeq(1.6\mbox{--}2.5)\times10^{-4}\,M_\odot$, where $\beta_{\rm sh}\equiv v_{\rm sh}/c$ is the shock velocity in the inner wind. Since the shock is decelerated after encountering the dense CSM shell, we expect $\beta_{\rm sh}\gtrsim\beta_{\rm bo}$. The value obtained by adopting $R_{\rm bo}=2\times10^{13}\ {\rm cm}$ and $\beta_{\rm bo}=0.05$--$0.08$ is therefore a conservative upper limit on the minimum wind mass. Even this upper limit is far smaller than the CSM shell mass inferred from the optical cooling emission of EP260321a, $M_{\rm CSM}\simeq0.06\,M_\odot$, and such an inner wind would therefore have a negligible effect on our calculation of the cooling emission.}

{
EP250108a is roughly an order of magnitude more  energetic in X-rays than EP260321a. Its prompt emission is difficult to reconcile with a standard shock breakout driven by ordinary SN ejecta \citep{EylesFerris2025,Srinivasaragavan2025}. The proposed interpretations include a relativistic shock breakout or a weak, successfully emerged jet \citep{Li2025}, a trapped or low-energy jet-driven explosion \citep{EylesFerris2025,Rastinejad2025}, and the breakout of a choked-jet cocoon from extended CSM \citep{Srinivasaragavan2025}.
EP250827b is more similar to EP260321a. Its duration, $T_{90}\simeq1200\ {\rm s}$, and X-ray energy, $E_{\rm X,iso}\sim10^{48}\ {\rm erg}$, are comparable to those of EP260321a. 
Its very soft spectrum could be fitted equally well by a power law with $\Gamma_{\rm X}=3.3_{-1.0}^{+1.7}$ or a blackbody with $kT=0.22\pm0.03\ {\rm keV}$, while a broken-power-law fit gives $E_{\rm p}<1.5\ {\rm keV}$. 
Its X-ray light curve appears to contain two broad pulses, whereas EP260321a shows a single smooth pulse. This more complex temporal structure is less characteristic of a simple spherical SBO, although it does not exclude breakout from a structured CSM. The first optical peak of SN~2025wkm was approximately an order of magnitude brighter than the sub-day optical component of SN~2026gzf.  \citet{Srinivasaragavan2026} suggested that the prompt X-rays are produced as fast engine-driven outflows (i.e., magnetically driven winds from the magnetar and the disk) break out through extended CSM. Shock-cooling emission from the CSM cannot account for the luminous first optical peak; the peak is instead modeled as photospheric cooling emission from adiabatically expanding  disk outflows and magnetar winds. This differs from EP260321a, for which the same dense CSM shell can account for both thermal X-ray breakout and early optical cooling emission.}

\begin{acknowledgments}
This work is supported by the
National Natural Science Foundation of China (grant
Nos. 12333006 and 12121003), the
Fundamental Research Funds for the Central Universities (KG202502), and the Nanjing University PhD Student Zhujian Program. We are grateful to the High Performance Computing Center of
Nanjing University for doing the numerical calculations in this
paper on its blade cluster system.
\end{acknowledgments}

\clearpage

\appendix

\section{Radio Afterglow Model}
\label{app:radio_model}

We calculate the radio emission from the forward shock driven by the post-breakout ejecta. After shock breakout, the fastest ejecta interact with the external medium, and the radiation-mediated shock transitions into a collisionless shock that accelerates relativistic electrons and produces synchrotron emission. As the blast wave decelerates, slower and more energetic ejecta catch up with it and continuously inject energy, as expected for shock-breakout ejecta with a steep energy--velocity profile \citep{Tan2001,Nakar2012,Duran2015}. The ejecta are described by a cumulative energy distribution
\begin{equation}
    E(>\beta)
    =
    E_{\rm f,0}
    \left(
        \frac{\beta}{\beta_{\rm f,0}}
    \right)^{-q_{\rm ej}},
    \label{eq:ejecta_energy_distribution}
\end{equation}
where $E_{\rm f,0}$ and $\beta_{\rm f,0}$ are the energy and dimensionless velocity of the fastest ejecta shell. 
Following Sect.~\ref{subsec:radio_velocity}, 
$q_{\rm ej}=s'-1=3.35+5.35/n$.
For a radiative envelope, shock-breakout ejecta have a steep outer profile with $q_{\rm ej}\simeq5$--6 \citep{Matzner1999,Tan2001,Duran2015}. 

For an external density profile $\rho_{\rm CSM}\propto r^{-k}$
($k=2$ for a steady wind), after the fastest shell decelerates,
the slower ejecta carrying most of the kinetic energy catch up with it and refresh the forward shock. The shocked energy then evolves as \citep{Duran2015}
\begin{equation}
\label{Eq_Esh_radio}
E_{\rm sh}(t)
=
E_{\rm f,0}
\left(
\frac{t}{t_{\rm dec}}
\right)^{
\frac{(3-k)(s'-1)}{4-k+s'}},
\qquad
t>t_{\rm dec},
\end{equation}
where $t_{\rm dec}$ is the deceleration time of the fastest ejecta; before deceleration, $E_{\rm sh}=E_{\rm f,0}$. 
For a wind medium, this reduces to $E_{\rm sh}\propto t^{(s'-1)/(s'+2)}$. 
The forward-shock synchrotron emission is set by the post-shock energy fractions in nonthermal electrons and magnetic fields, $\epsilon_e$ and $\epsilon_B$, and by the electron index $p$.

At radius $R$, the swept-up mass is
\begin{equation}
    M_{\rm sw}(R)
    =
    4\pi
    \int^R \rho_{\rm CSM}(r)r^2dr .
\end{equation}
The blast-wave dynamics are approximated by
\begin{equation}
    E_{\rm sh}
    \simeq
    (\Gamma_{\rm sh}\beta)^2M_{\rm sw}c^2 ,
    \label{eq:blastwave_energy}
\end{equation}
where $\Gamma_{\rm sh}=(1-\beta^2)^{-1/2}$ is the blast-wave Lorentz factor. 
This relation accounts for the kinetic and internal energies of the shocked material to order unity and reduces to $E_{\rm sh}\simeq M_{\rm sw}\beta^2c^2$ in the Newtonian limit.

The post-shock internal energy density is estimated as
\begin{equation}
e_{\rm sh}
\simeq
4\Gamma_{\rm sh}(\Gamma_{\rm sh}-1)\rho_{\rm CSM} c^2 .
\label{eq:esh}
\end{equation}
We parameterize the partition of this energy between magnetic fields and nonthermal electrons by $\epsilon_B$ and $\epsilon_e$, respectively. The downstream magnetic field is therefore
\begin{equation}
B=
\left(8\pi\epsilon_B e_{\rm sh}\right)^{1/2}.
\label{eq:Bfield}
\end{equation}
The accelerated electrons are assumed to follow a power-law distribution,
\begin{equation}
\frac{dN_e}{d\gamma_e}
\propto
\gamma_e^{-p},
\qquad
\gamma_e\ge\gamma_m .
\label{eq:electron_distribution}
\end{equation}
For relativistic electrons, the minimum Lorentz factor is set by the electron energy fraction,
\begin{equation}
\gamma_m-1
=
\frac{p-2}{p-1}
\epsilon_e
\frac{m_p}{m_e}
(\Gamma_{\rm sh}-1).
\label{eq:gammam}
\end{equation}
In the deep-Newtonian regime, we instead describe the accelerated electrons by a power law in momentum rather than in Lorentz factor \citep{Sironi2013}.

The synchrotron spectrum is specified by the injection, cooling, and self-absorption frequencies, $\nu_m$, $\nu_c$, and $\nu_a$. 
We compute the injection frequency from the downstream magnetic field and the minimum electron Lorentz factor, $\nu_m=(3q_eB/4\pi m_ec)\gamma_m^2$. 
The cooling Lorentz factor is $\gamma_c=6\pi m_ec/(\sigma_T B^2t')$, where $t'$ is the comoving dynamical time, giving $\nu_c=(3q_eB/4\pi m_ec)\gamma_c^2$. 
The self-absorption frequency is determined by $\tau_\nu(\nu_a)=1$. The observed flux is then constructed from the synchrotron spectrum, including synchrotron self-absorption \citep{Sari1998,Granot2002,Duran2015}. 
This framework allows us to assess how the initial breakout velocity and ejecta energy distribution affect the late-time radio luminosity.

\section{Cooling-emission Model}
\label{app:cooling_model}

We model the early optical data as cooling emission from the shocked CSM \citep[e.g.][]{NakarPiro2014,Piro2015}. Following these works, we assume that the shocked CSM is a uniform spherical shell that begins expanding at the breakout radius
$R_{\rm bo}$ with velocity $v_{\rm bo}$.
The shell expands with a constant velocity and cools adiabatically, so its radius evolves as
\begin{equation}
R(t)
=
R_{\rm bo}+v_{\rm bo}t
=
R_{\rm bo}\Lambda(t),
\qquad
\Lambda(t)=1+\frac{t}{t_{\rm exp}},
\end{equation}
where
\begin{equation}
t_{\rm exp}=\frac{R_{\rm bo}}{v_{\rm bo}} .
\end{equation}
The initial radius is retained because the diffusion radius of the shocked CSM is comparable to
$R_{\rm bo}$.
This differs from the usual cooling emission following shock breakout from the stellar surface, for which the initial radius is often much smaller than the radius at the optical peak \citep[e.g.][]{Zheng2025ApJ...985...21Z,Zhu2025}.

For a CSM shell of mass $M_{\rm CSM}$ and pre-shock width
$\Delta R_{\rm CSM}=f_\Delta R_{\rm bo}$, the shocked shell has an initial width
\begin{equation}
\Delta R_{\rm sh,0}
\simeq
\frac{\Delta R_{\rm CSM}}{\chi},
\end{equation}
where $\chi\simeq7$ is the compression ratio. In homologous expansion, the shell width evolves as
\begin{equation}
\Delta R_{\rm sh}(t)
=
\Delta R_{\rm sh,0}\Lambda(t).
\end{equation}
The optical depth and diffusion time are then
\begin{equation}
\tau(t)
\simeq
\frac{\kappa M_{\rm CSM}}
{4\pi R^{2}(t)},
\qquad
t_{\rm diff}(t)
\simeq
\frac{\tau(t)\Delta R_{\rm sh}(t)}
{c}
=
\frac{t_{d,0}}
{\Lambda(t)},
\end{equation}
where the initial diffusion time is
\begin{equation}
t_{d,0}
=
\frac{\kappa M_{\rm CSM}\Delta R_{\rm sh,0}}
{4\pi cR_{\rm bo}^{2}}
=
\frac{\kappa f_\Delta M_{\rm CSM}}
{4\pi\chi cR_{\rm bo}} .
\end{equation}
For a uniform pre-shock shell, this can also be written as
\begin{equation}
t_{d,0}
=
\frac{\kappa\rho_0R_{\rm bo}^{2}f_\Delta^{2}}
{\chi c}.
\end{equation}

The internal energy of the shocked CSM evolves as
\begin{equation}
\frac{dE}{dt}
=
-\frac{E}{t+t_{\rm exp}}
-
L_{\rm cool},
\qquad
L_{\rm cool}
=
\frac{E}{t_{\rm diff}(t)} ,
\end{equation}
where the first term describes adiabatic cooling and the second term describes radiative diffusion.
Solving this equation gives
\begin{equation}
E(t)
=
\frac{E_{\rm int,0}}
{\Lambda(t)}
\exp\left[
-\frac{t+t^{2}/(2t_{\rm exp})}
{t_{d,0}}
\right],
\end{equation}
and therefore
\begin{equation}
L_{\rm cool}(t)
=
\frac{E(t)}
{t_{\rm diff}(t)}
=
\frac{E_{\rm int,0}}
{t_{d,0}}
\exp\left[
-\frac{t+t^{2}/(2t_{\rm exp})}
{t_{d,0}}
\right].
\label{Eq_app_Lcool}
\end{equation}
Here $E_{\rm int,0}$ is the initial internal energy of the shocked CSM.


When the cooling emission is approximately thermalized, the bolometric luminosity defines the effective temperature,
\begin{equation}
T_{\rm eff}(t)
=
\left[
\frac{L_{\rm cool}(t)}
{4\pi\sigma_{\rm SB}R^{2}(t)}
\right]^{1/4}.
\end{equation}
To allow for a possible color correction, we introduce the color temperature $T_{\rm col}(t)=f_{\rm col}T_{\rm eff}(t)$ and compute the spectral luminosity as a diluted blackbody, $L_\nu(t)=4\pi^2 R^2(t)f_{\rm col}^{-4}B_\nu[T_{\rm col}(t)]$. Here $f_{\rm col}\sim1$ is the color-correction factor, and the factor $f_{\rm col}^{-4}$ preserves the bolometric luminosity, $\int L_\nu d\nu=L_{\rm cool}$.

A characteristic bolometric release time is obtained from
$t_{\rm diff}(t_{\rm p})\simeq t_{\rm p}$.
Using
$t_{\rm diff}(t)=t_{d,0}/(1+t/t_{\rm exp})$, this gives
\begin{equation}
t_{\rm p}
=
\frac{
-t_{\rm exp}
+
\left(t_{\rm exp}^{2}+4t_{d,0}t_{\rm exp}\right)^{1/2}
}{2}.
\end{equation}
Equivalently, defining
\begin{equation}
q\equiv\frac{t_{d,0}}{t_{\rm exp}},
\qquad
\Xi(q)
\equiv
\frac{(1+4q)^{1/2}-1}{2q},
\end{equation}
we have
\begin{equation}
t_{\rm p}
=
\Xi(q)t_{d,0}.
\end{equation}
Substituting this time into Eq.~\ref{Eq_app_Lcool} gives the bolometric luminosity at the characteristic release time,
\begin{equation}
L_{\rm p}
=
\frac{E_{\rm int,0}}
{t_{d,0}}
\exp\left[
-\Xi(q)-\frac{q\Xi^{2}(q)}{2}
\right].
\end{equation}
In Sect.~\ref{sec:csm_cooling}, we estimate the initial internal energy as
$E_{\rm int,0}\simeq M_{\rm CSM}v_{\rm bo}^{2}/2$.
For the fiducial marginally thermalized solution with $v_{\rm bo}\simeq0.05c$ (see Sect.~\ref{sec:shock_velocity}), we obtain $q\simeq1.9$ and $\Xi(q)\simeq0.51$, yielding $t_{\rm p}\simeq0.20\ {\rm d}$.
The corresponding exponential factor is $\simeq0.47$.

The optical-band maximum occurs later than the bolometric release time because the emitting area grows while the cooling spectrum shifts into the optical band.
For a blackbody spectrum, the spectral luminosity at frequency $\nu$ is
\begin{equation}
L_\nu(t)
=
4\pi^{2}R^{2}(t)B_\nu[T_{\rm eff}(t)] .
\end{equation}
Using
\begin{equation}
L_{\rm cool}(t)
=
\frac{E_{\rm int,0}}
{t_{d,0}}
\exp[-\zeta(\Lambda^{2}-1)],
\qquad
\zeta
\equiv
\frac{t_{\rm exp}}
{2t_{d,0}},
\end{equation}
the Rayleigh--Jeans limit gives
\begin{equation}
L_\nu(t)
\simeq
\frac{8\pi^{2}k_{\rm B}\nu^{2}}
{c^{2}}
R_{\rm bo}^{3/2}
\left(
\frac{E_{\rm int,0}}
{4\pi\sigma_{\rm SB}t_{d,0}}
\right)^{1/4}
\Lambda^{3/2}
\exp\left[
-\frac{\zeta(\Lambda^{2}-1)}{4}
\right].
\end{equation}
The peak time at frequency $\nu$, denoted by $t_{\nu,p}$, is obtained from
$d\ln L_\nu/dt=0$, or equivalently $d\ln L_\nu/d\Lambda=0$.
This gives
\begin{equation}
\Lambda_{\nu,p}
=
\left(
\frac{6t_{d,0}}
{t_{\rm exp}}
\right)^{1/2},
\qquad
t_{\nu,p}
=
t_{\rm exp}
\left[
\left(
\frac{6t_{d,0}}
{t_{\rm exp}}
\right)^{1/2}
-1
\right].
\end{equation}
In the Rayleigh--Jeans limit, the peak time is independent of frequency because the frequency dependence enters only through the multiplicative factor
$L_\nu\propto\nu^2$.

The corresponding Rayleigh--Jeans peak spectral luminosity is
\begin{align}
L_{\nu,p}^{\rm RJ}
&=
\frac{8\pi^{2}k_{\rm B}\nu^{2}}
{c^{2}}
R_{\rm bo}^{3/2}
\left(
\frac{E_{\rm int,0}}
{4\pi\sigma_{\rm SB}t_{d,0}}
\right)^{1/4}
\Lambda_{\nu,p}^{3/2}
\exp\left[
-\frac{\zeta(\Lambda_{\nu,p}^{2}-1)}{4}
\right].
\end{align}
For the fiducial marginally thermalized solution, the full expression gives an optical peak time of order
$t_{\nu,p}\sim0.5$--$0.7\ {\rm d}$.
The corresponding peak spectral luminosity is
\begin{equation}
L_{\nu,p}^{\rm RJ}
\sim
{\rm few}\times10^{27}
\left(
\frac{\nu}
{6\times10^{14}\ {\rm Hz}}
\right)^{2}
\ {\rm erg\ s^{-1}\ Hz^{-1}},
\end{equation}
or
$\nu L_{\nu,p}\sim10^{42}\ {\rm erg\ s^{-1}}$
in the optical band.
Thus, the shocked CSM naturally releases its trapped radiation on a sub-day timescale and can account for the bright early optical emission of EP260321a.

\clearpage

\bibliography{sample702}{}
\bibliographystyle{aasjournalv7.1}

\end{document}